\documentclass[journal]{IEEEtran}

\usepackage{amsmath,amsfonts}
\usepackage{algorithmic}
\usepackage{algorithm}
\usepackage{graphicx}

\usepackage{array}
\usepackage{textcomp}
\usepackage{url}
\usepackage{verbatim}
\usepackage{graphicx,color,xcolor}
\usepackage{multirow}
\usepackage[colorinlistoftodos]{todonotes}
\usepackage{booktabs}
\usepackage{fdsymbol}
\usepackage[export]{adjustbox}
\ifCLASSOPTIONcompsoc
    \usepackage[caption=false, font=normalsize, labelfont=sf, textfont=sf]{subfig}
\else
\usepackage[caption=false, font=footnotesize]{subfig}
\fi

\usepackage{fixltx2e}
\usepackage{url}
\usepackage{svg}

\begin{document}

\title{Generalizing Speaker Verification for Spoof Awareness in the Embedding Space}
%
%
%

\author{Xuechen~Liu,
        Md~Sahidullah,~\IEEEmembership{Member,~IEEE,}
        Kong~Aik~Lee,~\IEEEmembership{Senior Member,~IEEE,}
        and~Tomi~Kinnunen
\thanks{Xuechen Liu is with Universit\'{e} de Lorraine, CNRS, Inria, LORIA, F-54000, Nancy, France and  the School of Computing, University of Eastern Finland, FI-80101, Joensuu, Finland. He is now with National Institute of Informatics, Tokyo 101-8430, Japan (e-mail: xuecliu@nii.ac.jp).}
\thanks{Md Sahidullah is with Institute for Advancing Intelligence, TCG CREST, Kolkata, West Bengal 700091, India (e-mail: md.sahidullah@tcgcrest.org).}
\thanks{Kong Aik Lee is with the Department of Electrical and Electronic Engineering, The Hong Kong Polytechnic University, Hong Kong (e-mail: kong-aik.lee@polyu.edu.hk).}
\thanks{Tomi Kinnunen is with School of Computing, University of Eastern Finland, FI-80101, Joensuu, Finland (e-mail: tomi.kinnunen@uef.fi).}
\thanks{Published in IEEE/ACM-TASLP (doi: 10.1109/TASLP.2024.3358056) \copyright 2024 IEEE. Personal use of this material is permitted. Permission from IEEE must be obtained for all other uses, in any current or future media, including reprinting/republishing this material for advertising or promotional purposes, creating new collective works, for resale or redistribution to servers or lists, or reuse of any copyrighted component of this work in other works.}
}

\markboth{PUBLISHED IN IEEE/ACM TRANSACTIONS ON AUDIO, SPEECH, AND LANGUAGE PROCESSING (10.1109/TASLP.2024.3358056)}%
{Shell \MakeLowercase{\textit{et al.}}: Bare Demo of IEEEtran.cls for IEEE Journals}

\maketitle

\begin{abstract} 
It is now well-known that \emph{automatic speaker verification} (ASV) systems can be spoofed using various types of adversaries. The usual approach to counteract ASV systems against such attacks is to develop a separate spoofing \emph{countermeasure} (CM) module to classify speech input either as a bonafide, or a spoofed utterance. Nevertheless, such a design requires additional computation and utilization efforts at the authentication stage. An alternative strategy involves a single monolithic ASV system designed to handle both zero-effort imposter (non-targets) and spoofing attacks. Such \emph{spoof-aware} ASV systems have the potential to provide stronger protections and more economic computations. To this end, we propose to generalize the standalone ASV (G-SASV) against spoofing attacks, where we leverage limited training data from CM to enhance a simple backend in the embedding space, without the involvement of a separate CM module during the test (authentication) phase. We propose a novel yet simple backend classifier based on deep neural networks and conduct the study via domain adaptation and multi-task integration of spoof embeddings at the training stage. Experiments are conducted on the ASVspoof 2019 logical access dataset, where we improve the performance of statistical ASV backends on the joint (bonafide and spoofed) and spoofed conditions by a maximum of 36.2\% and 49.8\% in terms of equal error rates, respectively.
\end{abstract}


%
\IEEEpeerreviewmaketitle

\section{Introduction}
\label{sec:intro}
\IEEEPARstart{A}{utomatic} speaker verification (ASV) \cite{asv-dnn-review_2021} is widely regarded as a convenient approach for biometric authentication. Starting with early methods based on statistical models such as \emph{joint factor analysis} \cite{jfa2008} and \emph{i-vectors} \cite{ivector2010}, modern ASV systems \cite{xvector2018, etdnn, ecapa_tdnn2020} are largely based on deep neural nets (DNNs). 
Despite its effectiveness \cite{asv-dnn-review_2021}, ASV performance is known to degrade substantially in the presence of fabricated or manipulated speech inputs \cite{wu2015spoofing}. These \emph{spoofing attacks} \cite{asvspoof2015}, or \emph{representation attacks}\footnote{\url{https://www.iso.org/standard/53227.html}} can be constructed using text-to-speech (TTS) methods, voice conversion, and speech replay~\cite{wu2015spoofing,voicepad2019}. The vulnerability of ASV systems has been reported by the automatic speaker verification spoofing and countermeasures (ASVspoof)~\cite{asvspoof2015,kinnunen17_interspeech,todisco19_interspeech,yamagishi21_asvspoof} challenge and other spoofed speech databases~\cite{add2022, partialspoof2022}.

In order to improve the robustness of the ASV system against spoofing attacks, \emph{countermeasures} (CM) are developed. The function of a CM is to detect computer-generated or replayed speech samples~\cite{wu2015spoofing,voicepad2019}. Driven by the ASVspoof initiative~\cite{asvspoof2015}, a number of CMs have been developed to detect specific types of spoofing attacks. The most common approach is to first design a standalone CM which is subsequently integrated with a conventional ASV system. The CM and ASV can be combined in a cascade \cite{Kinnunen2020-tdcf-fundamentals} or parallel fashion~\cite{sahidullah16_interspeech, todisco19_interspeech}. This integration can take place at the score~\cite{sahidullah16_interspeech}, decision~\cite{sahidullah16_interspeech}, or embedding levels~\cite{prob_fusion_asvspoof2022}. 

Compared to an unprotected ASV system, the combination of a CM and an ASV subsystem has shown substantial performance improvement for ASV in the face of spoofing attacks \cite{aasist2022, ssl_frontends2022, rawboost2022, prob_fusion_asvspoof2022}. Nonetheless, this approach is not entirely without problems. First and foremost, developing an independent CM requires additional design effort.
Since anti-spoofing is a different task from speaker verification, 
additional computation and utilization efforts will be required at the authentication stage. For instance, some CM modules usually require their specialized feature extractors (e.g. \cite{cqcc_max, spoof_features2015}) and classifiers (e.g. \cite{lcnn}). 
Even though the recent approaches can make use of spectrogram \cite{spoof_cnn_attentive2019} and raw audio waveform \cite{aasist2022, spoof_cnn_e2e2021, samo2023}, this typically results in
increased
computational demands and resource utilization, rendering the system less suitable for specific scenarios, such as handheld devices.
In addition to retaining low computational costs, one should also ensure that both the ASV and CM \emph{generalize} to unseen test conditions; otherwise, the performance of the combined system (often addressed using score fusion or cascaded approach \cite{Kinnunen2020-tdcf-fundamentals}) 
may become dependent on whichever of the two systems is weaker\footnote{Here we refer to a subsystem that experiences larger relative degradation in the discrimination performance between training and evaluation. In our context, typically this is the spoof countermeasure \cite{asvspoof2015, asvspoof2019}.}.
While modern ASV systems generalize fairly well, the performance of CM systems is highly dependent on external factors such as noise, channel, etc. \cite{dku_cmri_asvspoof2021, spoof_detection_noisy2016}.

Driven by these motivations, a number of first steps towards \emph{single} ASV systems capable of handling both conventional (bonafide) and spoofed speech inputs have been taken. While the early work on such \emph{spoof-aware} ASV \cite{joint_opt_ivector2015} considered i-vector speaker embeddings (with a dedicated statistical backend), recent studies \cite{joint_opt_sasv2020} have focused on neural embeddings generated by ASV and/or CM\footnote{As opposed to \emph{speaker} embedding, the latter might be called \emph{spoof} embedding.}. A representative example of such recent developments can be found in the baseline used in the recent \emph{spoof-aware ASV} (SASV) challenge \cite{prob_fusion_asvspoof2022}, where a fusion system was developed with three embeddings as the input: speaker embeddings from enrollment and test samples, and a spoofing embedding from the test sample. Similar embedding fusion approaches were also widely adopted by the SASV challenge participants (e.g.~\cite{sasv_embedding2022, prob_fusion_asvspoof2022}).

While the initial steps have been taken toward designing spoof-aware ASV, these investigations remain in their infancy. The new approach adopted in this study is based on a neural network-based backend system that uses only \emph{two} embeddings at the authentication stage: the enrollment and test (speaker) embeddings. We propose such an approach to address the following three primary problems:
\begin{itemize}
    \item \textbf{Removal of the CM from the authentication stage}. This has been marginally addressed in~\cite{sasv_personal_odyssey2022} with a conventional classifier based on \emph{probabilistic linear discriminant analysis} (PLDA). We replace PLDA, which is a linear-Gaussian model, with a simple neural network that can effectively model the spoofing ingredients in a supervised manner, with improved system performance.
    \item \textbf{Training with limited data}. Unlike common ASV systems trained on massive datasets, standardized spoof data is difficult to obtain in large quantities. This calls for new methods for the effective utilization of limited (spoofed) training datasets.
    \item \textbf{Integrating the latent space described by the spoof embeddings via \emph{multi-task learning} \cite{mtl1997}}. This allows training in an embedding space where spoof features are not necessary at authentication.
\end{itemize}


Motivated by the above concerns, in this study, we investigate the problem of generalizing an ASV system against the spoofing attack, to free the system from the presence of a standalone CM module at the authentication phase. We summarize the main contributions as follows:
\begin{itemize}
    \item We present a unified while scaled definition of various tasks involved in the spoofing scenario: \emph{spoof countermeasure}, \emph{joint optimization}, and \emph{spoof-aware speaker verification} (SASV). We then situate our work on \emph{generalizing a standalone ASV (G-SASV) system against the presentation attack}. Our task definition is escorted by data and a baseline system. The training and evaluation protocols are built based on publicly-available datasets in order for the research to be furthered. 
    
    \item To address the challenge of limited CM training data and reliance on standalone CM at authentication,
    we conduct preliminary attempts at generalizing the ASV backend by acquiring the spoofing information at the training stage only. We begin by applying domain adaptation (DA) techniques. We then model the latent embedding space by integrating the spoof features via multi-task learning with multiple extensions. Such ``spoof features" can take the form of ASV speaker embeddings (extracted from the ASV embedding extractor), CM embeddings (extracted from established CM pre-trained models), and additionally, attributes extracted from the available metadata. 
\end{itemize}

The application of domain adaptation techniques on spoof-aware speaker verification is addressed in \cite{sasv_personal_odyssey2022}. However, it is trained in an unsupervised fashion, and the performance is not at a usable level against the spoofing attacks. The application of multi-task learning under the spoofing scenario is also addressed in \cite{multitasking_antispoofing2019, multi_task_asvspoof2020}; However, it performs joint optimization and utilizes a CM component consistently, instead of empowering the ASV. Therefore, to the best of our knowledge, this is the first study on improving the generalization ability of a bare-bone ASV system against audio spoofing attacks, without the presence of CM in the authentication phase. 

\begin{figure}[th]
\centering
\subfloat[Spoof Countermeasure]{
\includegraphics[width=0.49\linewidth]{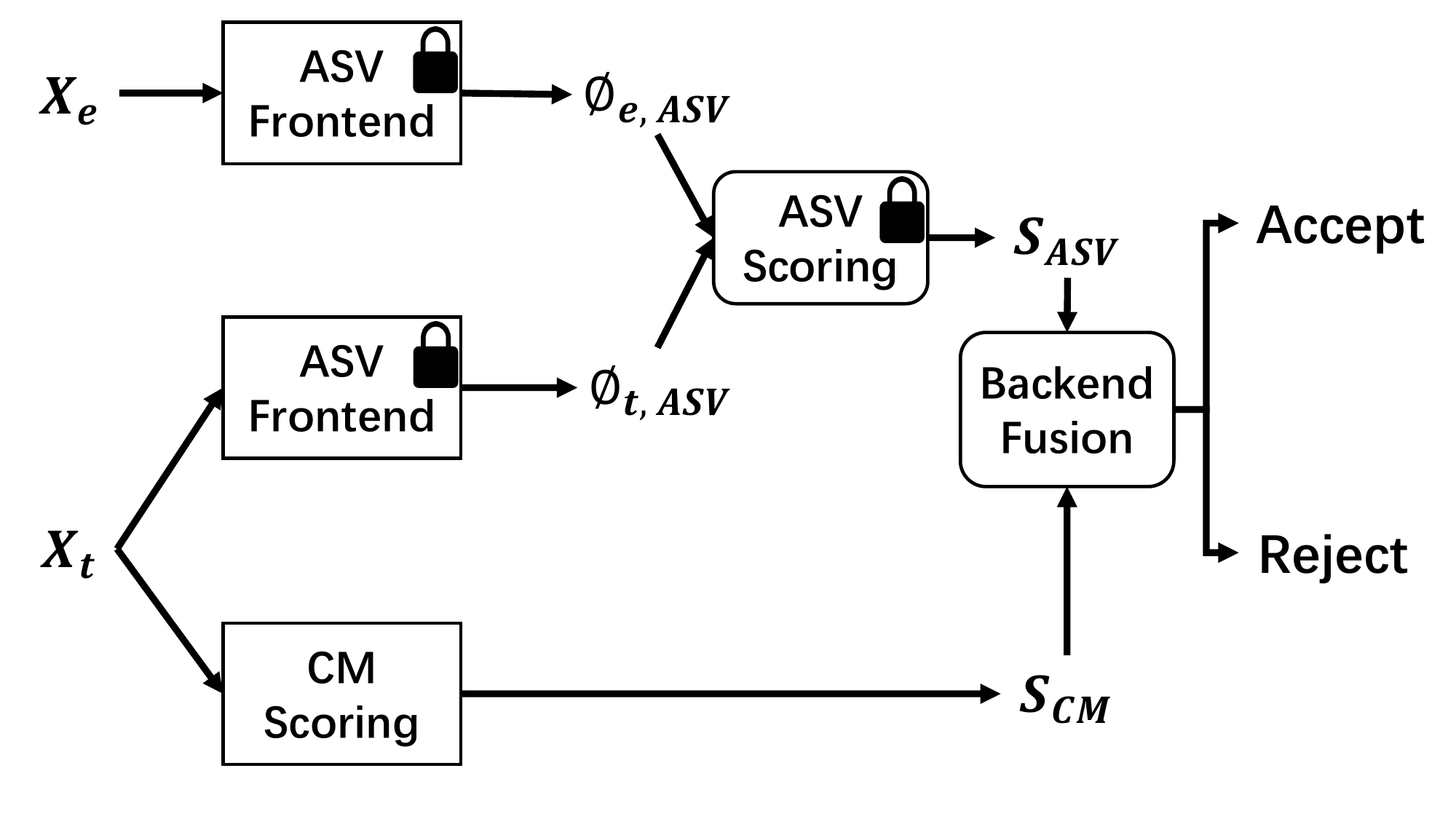} 
\label{fig7:antispoofing}
}
\subfloat[Joint Optimization]{
\includegraphics[width=0.49\linewidth]{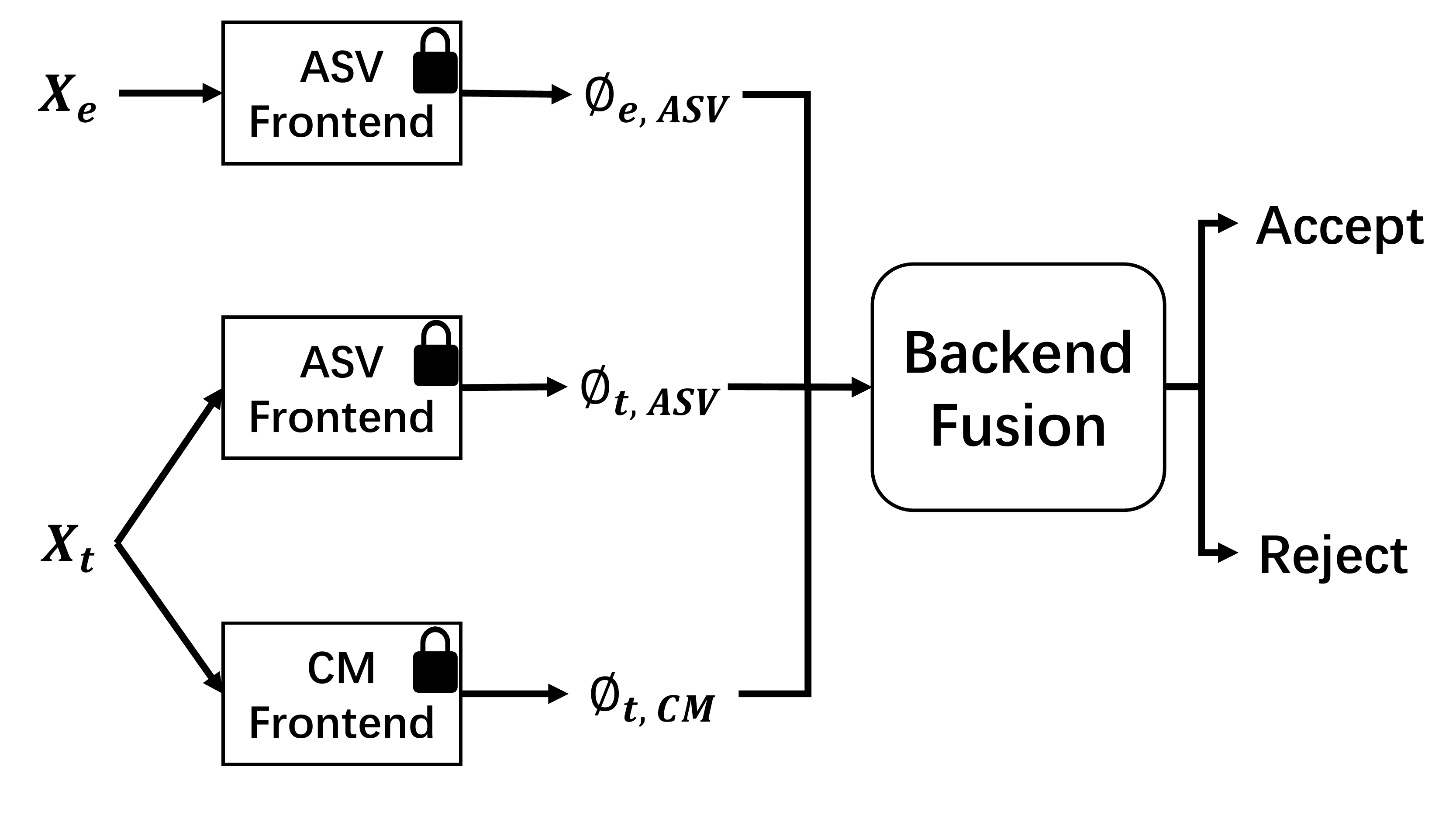} 
\label{fig7:jointop}
}

\subfloat[SASV]{
\includegraphics[width=0.49\linewidth]{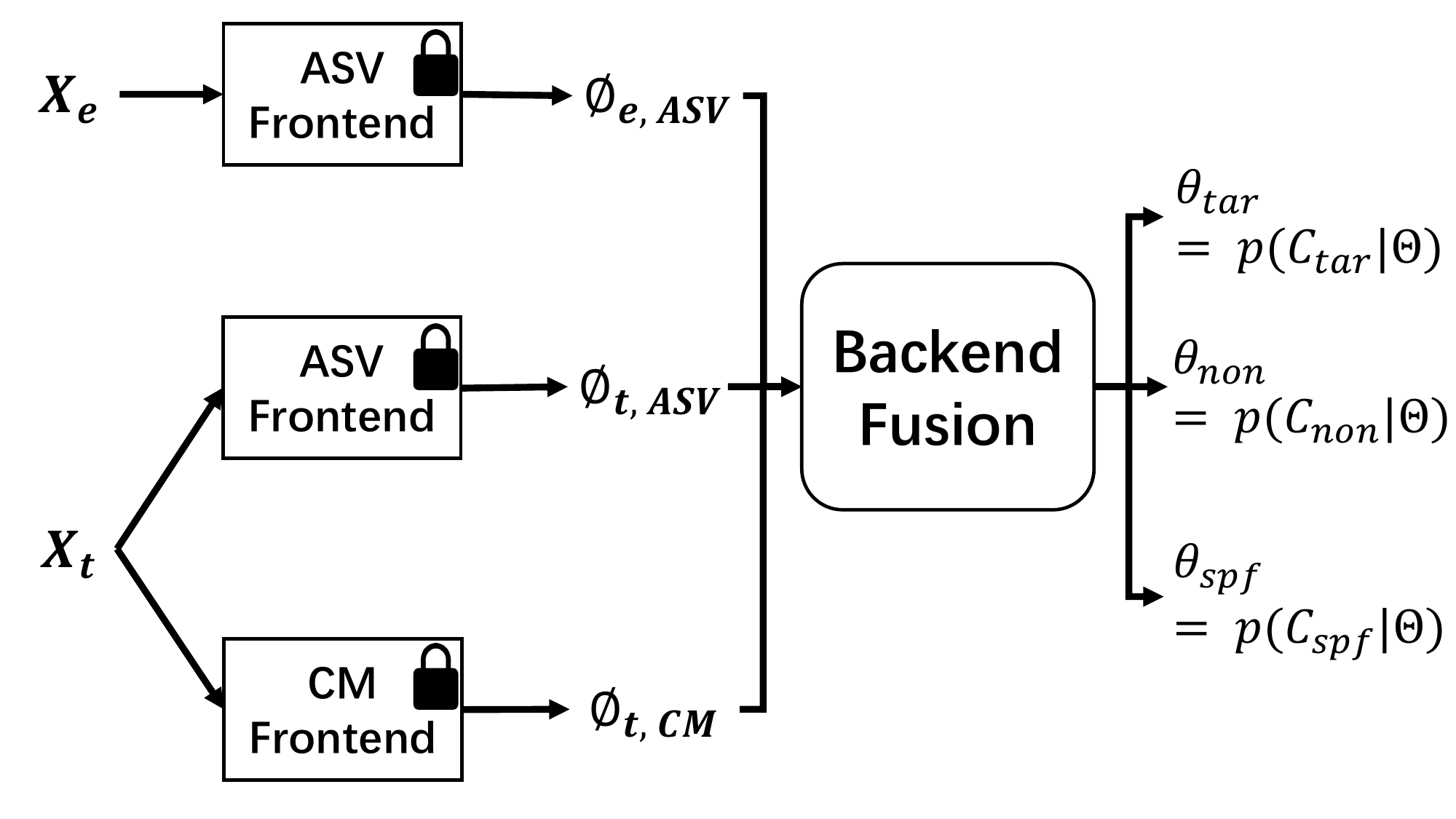} 
\label{fig7:sasv}
}
\subfloat[Generalized SASV]{
\includegraphics[width=0.49\linewidth]{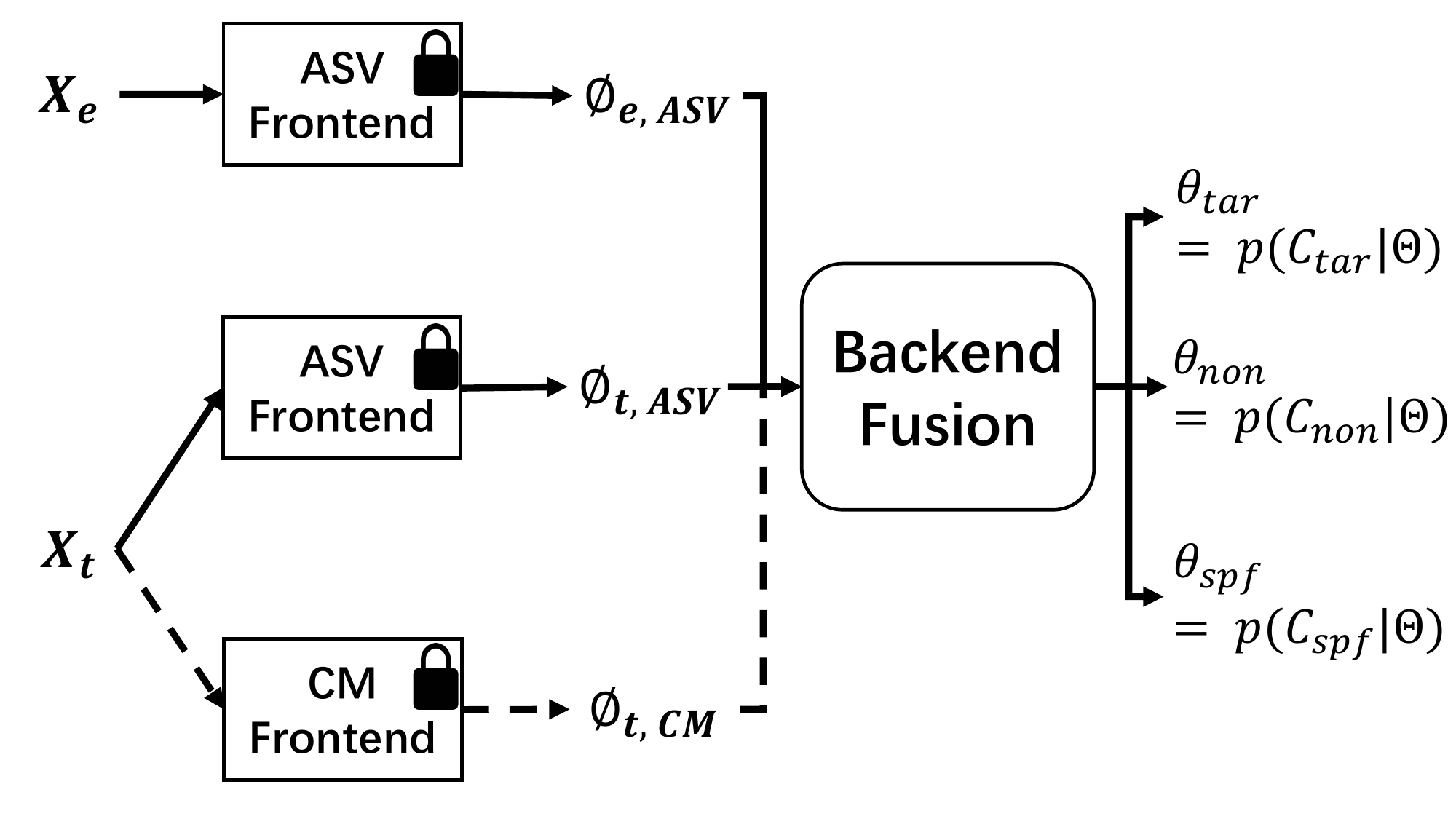} \label{fig7:gsasv}
}

\caption{A demonstration of existing approaches to improve spoofing robustness of ASV systems ((a) to (c)), with the proposed approach ((d)). Modules bearing ``Frontend" in their name extract embeddings, whereas modules labeled with ``scoring" provide decision scores based on the input(s). Modules featuring a lock symbol are sourced from pre-trained models and, as such, remain non-trainable during the training process.}
\label{fig:problem_definitions}
\end{figure}

\section{Problem Formulation}
In this section, we define the problem of SASV and discuss its connection to previous standalone spoof countermeasures, as well as the joint optimization of CM and ASV systems.

We begin by outlining the general functions of ASV and CM. First, we look at a typical \textbf{standalone ASV system} designed to process verification trials. Let us consider an ASV trial in the context of DNN-based modern ASV systems. The trial consists of an enrollment and test sample. The test sample normally corresponds to one utterance while the enrollment sample can either be built via single or multiple utterances. Their waveforms are $\mathcal{X}_\text{e}$ and $\mathcal{X}_\text{t}$, respectively. The respective speaker embeddings are computed by speaker embedding extractors based on deep neural networks such as ECAPA-TDNN \cite{ecapa_tdnn2020}, and are denoted by $\mathbf{\phi_\text{e,ASV}} = \Phi_{\Theta_\text{ASV}}(\mathcal{X}_\text{e})$ and $\mathbf{\phi_\text{t,ASV}} = \Phi_{\Theta_\text{ASV}}(\mathcal{X}_\text{t})$. This pair of embeddings, $(\mathbf{\phi_\text{e,ASV}} , \mathbf{\phi_\text{t,ASV}})$, is then forwarded to a scoring backend to obtain a scalar detection score of $s_\text{ASV} = \mathcal{S}_\text{ASV}(\mathbf{\phi_\text{e,ASV}} , \mathbf{\phi_\text{t,ASV}})$. Whenever a hard decision is required (as is the case for most operational implementations), the detection score is compared to a threshold of $\tau_\text{ASV}$, to make a binary prediction on whether the speakers of the two utterances are the same or different. Common choices for scoring include non-parametric cosine similarity \cite{cosine_similarity} or parametric, trainable models such as \emph{probabilistic linear discriminant analysis} (PLDA) \cite{plda_2006}.

We then consider a prototypical \textbf{standalone CM module}. Unlike an ASV system that provides classification decisions for an enrollment-test utterance pair, the input for a CM system is typically only the test utterance, $\mathcal{O}_{t}$~\cite{wu2015spoofing}. Each test utterance originates either from a bonafide human being or a spoofing attack. The task of the CM, then, is to predict which one of these two mutually exclusive hypotheses is true. Assuming the CM binary classifier is parameterized by $\Theta_\text{CM}$, the output is $\Phi_{\Theta_\text{CM}}(\mathcal{X}_\text{t})$, which can take various forms such as a score $s_\text{CM}$ \cite{asvspoof2015, wang2020asvspoof} or an embedding vector $\phi_\text{t,CM}$ \cite{aasist2022, prob_fusion_asvspoof2022}. Such output is then fed into a decision-making module where the score is compared to a CM threshold, indicating whether the input audio is spoofed or a bonafide human sample.

With the above key definitions outlined, we now present the definition of various solutions to tackle spoofing in the three tasks illustrated in Fig.~\ref{fig:problem_definitions}, along with a review of recent literature. As detailed below and depicted, the main difference between the three tasks is on: 1) The training target classes; 2) The necessity of standalone CM module at different stages.

\subsection{Spoof Countermeasure}
The main aim of spoof countermeasure, as illustrated in Fig. \ref{fig7:antispoofing},
is to distinguish between bonafide and spoofed biometric data, where the ASV system is fixed and the CM is trained. In this case, the fixed ASV system provides score $s_\text{ASV}$, while the output of the trained CM provides another score $s_\text{CM}$. The two scores are fused using a backend (for example, one from \cite{parallel_cm_max2018}) for decision-making with a threshold. The system can be placed either in a cascaded or parallel fashion. Fig. \ref{fig7:antispoofing} demonstrates the parallel design.

In response to the threat from the spoofing attack on the ASV system, a number of countermeasures have been developed. Moving on from conventional models based on \emph{Gaussian mixture models} and score-level integration with an ASV system \cite{sahidullah16_interspeech, parallel_cm_max2018}, DNN has been widely acquired in the design of CM, with a recent focus on convolutional architectures \cite{spoof_cnn_attentive2019, spoof_cnn_lightcnn2020, spoof_cnn_oneclass2021, lcnn} and end-to-end architectures \cite{rawnet2_spoofing2021, spoof_cnn_e2e2021}. Spectral information has been exploited in \cite{rawnet2_spoofing2021} and \cite{aasist2022}. Meanwhile, a number of advanced methods to augment the small-sized spoofing data have been developed \cite{rawboost2022, dataaug_spoofing2022}. In particular, the organizer of the ASVspoof2021 challenge \cite{yamagishi21_asvspoof} transmitted both bonafide and spoofed audios through a set of telephony systems including voice-over-IP (VoIP) and a public switched telephone network (PSTN). 

However, with above works noted, this binary classification task places its primary focus on designing a standalone CM module to assist the ASV system, which needs to be present at both training and authentication stages. 
The ASV system itself, if the CM is not present, remains vulnerable to spoofing attacks. Such vulnerability has been rarely addressed in the past. However, recently concerns have emerged, which are detailed next.

\subsection{Joint Optimization}
As illustrated in Fig. \ref{fig7:jointop}, given the ASV and CM information, the main focus of joint optimization task is to develop an integrated system that accepts inputs from the two classifiers and outputs the decision. The information provided by both ASV and CM systems can include either the three embeddings ($\mathbf{\phi_\text{e,ASV}}$, $\mathbf{\phi_\text{t,ASV}}$, $\mathbf{\phi_\text{t,CM}}$) or the two scores ($s_\text{ASV}$, $s_\text{CM}$). The presented backend generates the scores for the binary decision. 

The hypothesis thus is the same as the one from the spoof countermeasure. Fig. \ref{fig7:jointop} shows the example where the embeddings are used for the fusion backend. The key difference between this setup and the setup above is the CM system --- The primary focus of spoof countermeasure is to train a CM system, while the scoring backend is rather minimal. By contrast, here, we place our sole focus on the backend, with the CM model being fixed along with ASV.

Aiming to tackle the problem of the ASV system remaining problematic with the separate design of CM, there are notable works focusing on joint optimization of an ASV and CM module via a backend. Such optimization originated from jointly optimizing an i-vector model and a backend based on \emph{probabilistic linear discriminant analysis} (PLDA) \cite{joint_opt_ivector2015} and have recently been addressed via DNN architectures \cite{prob_fusion_asvspoof2022, joint_opt_sasv2020, jointop2022_stackfeats}. In \cite{joint_opt_sasv2020}, three types of embeddings --- ASV and the CM embeddings from both LA and PA scenarios --- were separately trained before being integrated into a fully-connected network with a softmax output layer. The output from the network is posterior probabilities with respect to three classes - bonafide target, bonafide non-target, and spoof target. When making the decision, the bonafide target is regarded as the positive class, while the latter two are regarded as the negative classes. A similar approach is proposed in \cite{prob_fusion_asvspoof2022}, where the posterior probabilities corresponding to the classes were integrated to make the decision via product rule \cite{product_rule}, and the output from the pre-trained models was concatenated at both the score-level and embedding-level, returning binary decisions. 
Recently, the multi-task learning of ASV and CM in a single setup was addressed in \cite{multi_task_asvspoof2020}, where the type of attack and speaker class are attached alongside the input acoustic features. Such joint optimization has been addressed in the recent \emph{spoof-aware speaker verification} challenge \cite{sasv2022}, which, unlike earlier ASVspoof-related works, places a strong emphasis on how to effectively integrate the CM module with pre-trained ASV models. The effect of stacking the outputs of ASV and CM as sub-modules is meanwhile addressed in \cite{jointop2022_stackfeats}.

The aforementioned works approached this task and reached reasonable performance on both speaker verification and anti-spoofing. Nevertheless, the task itself contains two problems. First, the training targets remain binary. While enhancing the performance on the anti-spoofing scenario, may be detrimental to the ASV performance, as shown in \cite{sasv2022}. Second, a standalone CM still should be present consistently at the training and authentication stages. The task described in the next section attempts to handle these two problems.

\begin{figure*}[t]
\centering
\subfloat[Conceptual framework\label{fig:sasv_framework}]{
\includegraphics[width=0.5\linewidth]{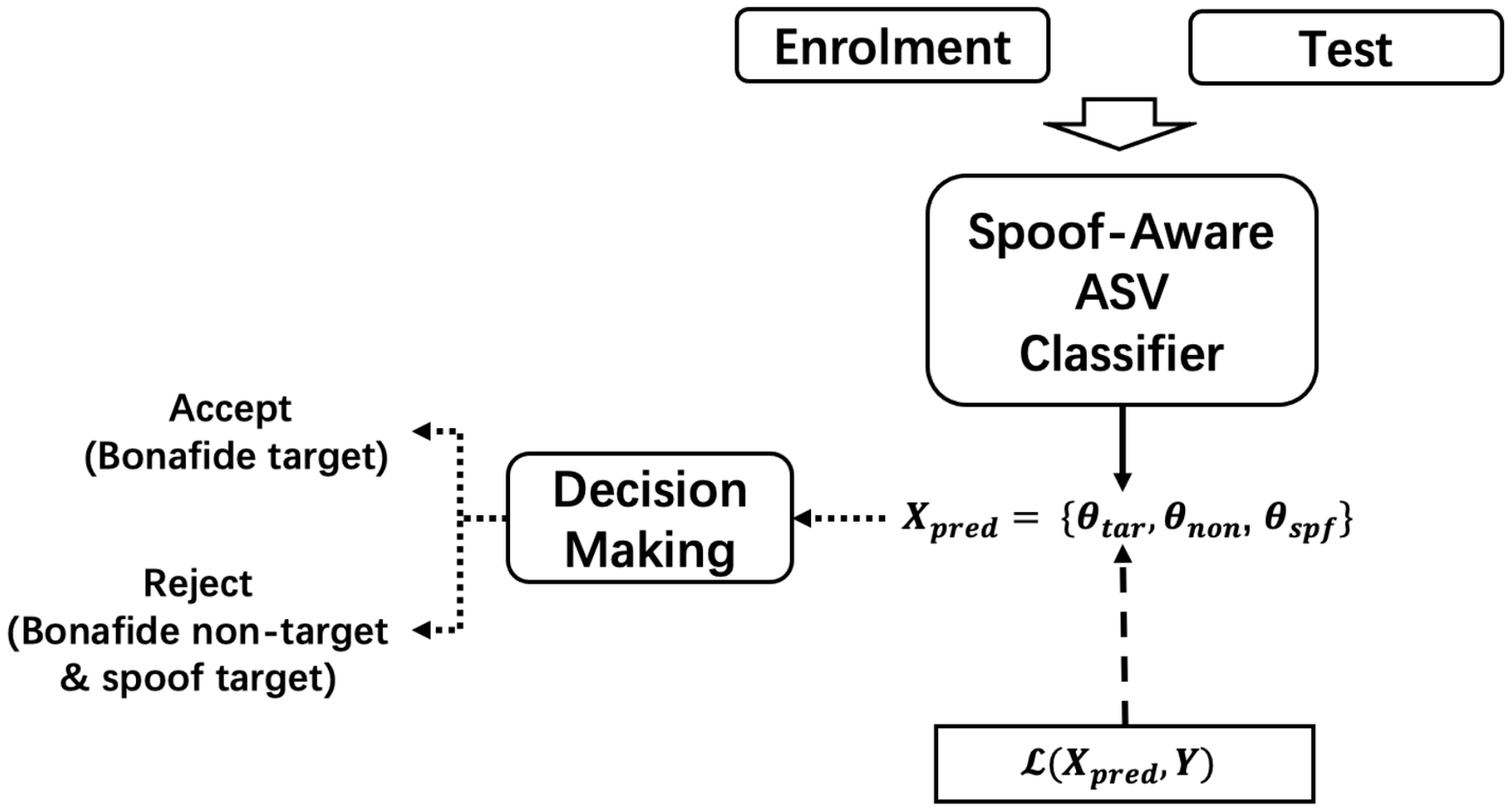} 
}
\subfloat[Baseline implementation\label{fig:baseline_protocol}]{
\includegraphics[width=0.5\linewidth]{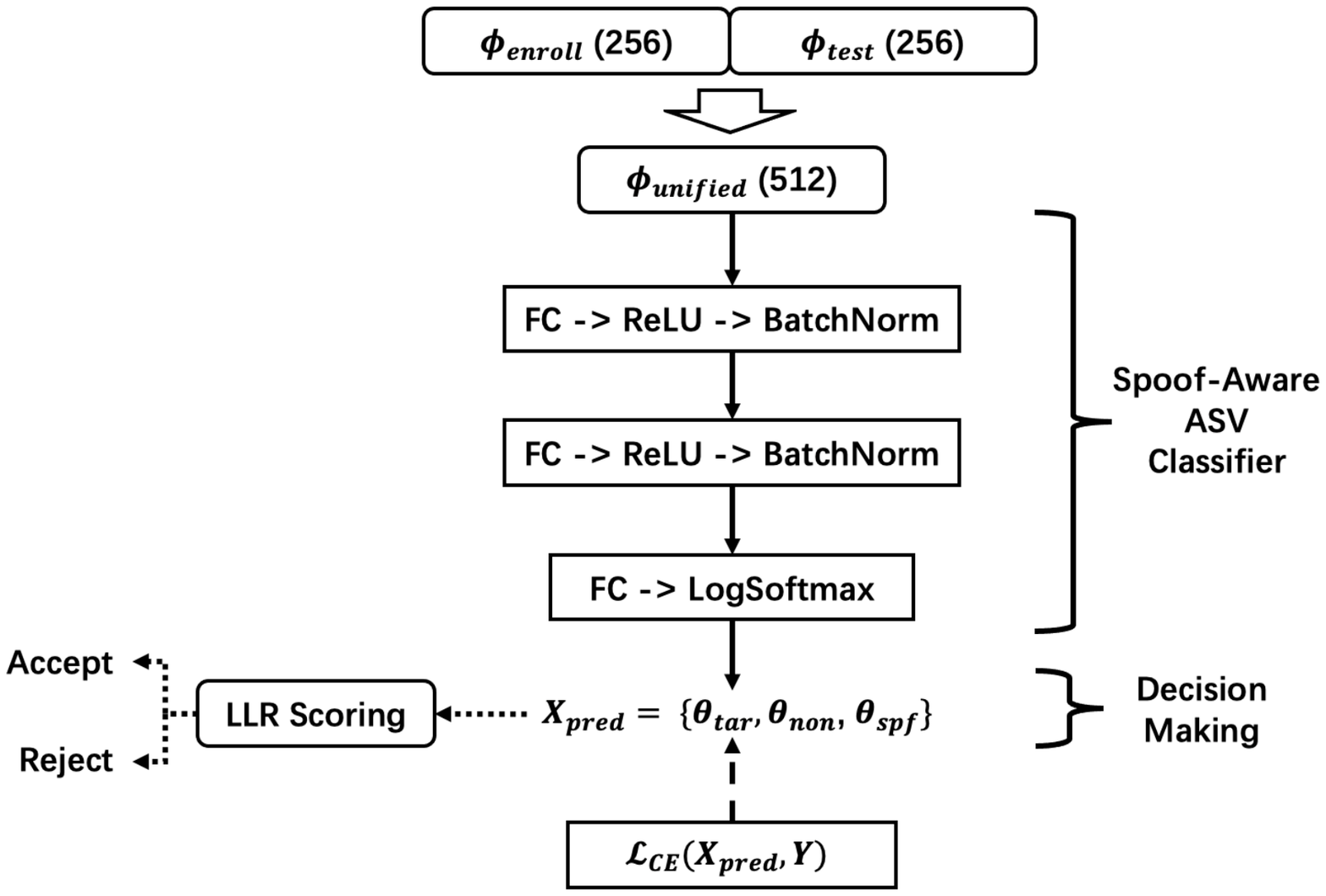} 
}
\caption{The framework of generalizing SASV against the spoofing attacks. The sub-figures demonstrates G-SASV at conceptual and practical level, respectively. The right-hand side outlines the baseline system. The three posterior probabilities represent the three classes respectively: bonafide target $C_\text{tar}$, bonafide non-target/impostor $C_\text{non}$, and spoof target $C_\text{spf}$. $\Theta$ denotes the set of parameters of the spoof-aware ASV classifier. $\mathcal{L}$ is the training loss function. The decision-making module is utilized for performing analysis analogously to anti-spoofing and joint optimization systems. The dashed line indicates the step which is discarded at the evaluation stage. The dotted lines indicate steps that are only executed during the evaluation stage.}
\end{figure*}

\subsection{Spoof-Aware Speaker Verification}
Analogous to the task above, for this task, a backend system is designed to integrate the information at the score-level or embedding-level. What differs from the above, as shown in Fig. \ref{fig7:sasv}, is the
training targets, which are three posterior probabilities: bonafide target $\theta_\text{tar}$, bonafide non-target $\theta_\text{non}$, and spoof target $\theta_\text{spf}$. The decision policy afterward, in this case, can be flexible --- it does not necessarily have to be the binary one in the above two problems (although in this study, for evaluation, we keep the binary decision as the final output, with a decision policy module).

Research effort on this task lies mainly in the effective ensemble of ASV and CM modules. The integration started from simple score-level fusion and furthered with embedding-level fusion \cite{prob_fusion_asvspoof2022}. The backend design has been furthered from there. For example, in \cite{sasv_attn_backend2022}, CM and ASV scores are acquired via parameterized nonlinear transformations based on the attention mechanism. A sub-network approach is presented in \cite{sasv_subnetwork2022}, where the intermediates of an ASV backbone network are used to generate a smaller network for spoofing detection. Scores output by the ASV system are also used for tuning CM scores, via transition to hard decisions through thresholding \cite{sasv_finetuingcm2022}. Ensembles of multiple CM models in multi-modal fashion are studied in \cite{prob_fusion_asvspoof2022}. End-to-end joint solutions with contrastive speaker verification loss and mixup \cite{mixup} training scheme are proposed in \cite{sasv_e2e2022}.

Nevertheless, all works mentioned above focus on either optimization of a CM system or the optimization of a backend with ASV and CM modules, which are constantly available in both training and evaluation. In this work, we rather focus on generalizing the ASV system itself via a backend. 
This was addressed in \cite{sasv_personal_odyssey2022} using a conventional PLDA backend \cite{kaldi_aplda} adapted with spoofed data in an unsupervised manner. In this work, we consider instead a simple neural network trained in a supervised manner to improve performance.
In such way, as indicated in Fig. \ref{fig7:sasv}, the presence of CM module can be flexible at authentication stage.

\section{Generalizing Speaker Verification}
\label{sec:sgsv_def}

In this section, we present the task of \emph{generalizing SASV} (G-SASV). The target of the "generalization power" of the system work on is against various types of spoofing attacks. As depicted in Fig. \ref{fig7:gsasv} along with earlier approaches, the main difference of this approach is that it eliminates the need for the CM at authentication stage.

The G-SASV approach is outlined in Fig. \ref{fig:sasv_framework}. The proposed classifier produces the according three posterior probabilities for bonafide target $\theta_\mathrm{tar}$, bonafide non-target $\theta_\mathrm{non}$, and spoof target $\theta_\mathrm{spf}$. Therefore, it involves a single classifier, instead of maintaining two different types of classifiers --- ASV and CM (and one of them with a low level of generalizability) at the same time. Based on this framework, there are three modules to be specified, which are explained below with our baseline implementation details. An illustration of the implementation is presented in Fig.~\ref{fig:baseline_protocol}.

\subsubsection{The source and type of the input}
In this study, without the essential presence of a pre-trained CM model, we use the concatenation of conventional speaker embeddings $\mathbf{\phi_\text{unified}} = (\mathbf{\phi_\text{enroll}}, \mathbf{\phi_\text{test}})$ as the source representation, which is extracted from a pre-trained ASV model. $\mathbf{\phi}_\text{enroll}$ and $\mathbf{\phi}_\text{test}$ correspond to $\mathbf{\phi_\text{e,ASV}}$ and $\mathbf{\phi_\text{t,ASV}}$ in Fig.~\ref{fig7:sasv}, respectively. This approach is a simplification of prior works, where the enrolment information is derived from a pre-trained ASV model, either at score-level \cite{wang2020asvspoof, parallel_cm_max2018} or embedding-level \cite{prob_fusion_asvspoof2022}. The test information, on the other hand, can be derived from either ASV or CM models, and also at both levels (e.g. \cite{joint_opt_sasv2020}).

\subsubsection{The spoof-aware ASV classifier}
The design of this module depends on both the input and output information provided. While there are numbers of established works maintaining advanced neural architectures as well as statistical ones (summarized in \cite{sasv2022, prob_fusion_asvspoof2022, yamagishi21_asvspoof}), as a preliminary study, we start from a simple 3-layer \emph{multi-layer perceptron} (MLP), as illustrated in Fig. \ref{fig:baseline_protocol}. Each of the first two layers is a cascade of affine operation and ReLU activation \cite{relu}, followed by batch normalization (BN) \cite{batchnorm}. The last layer is a cascade of affine transform and log-softmax \cite{logsoftmax}. The output is the softmax-normalized posterior probabilities concerning the three classes. 
Denoting number of classes as $C = 3$, the network is by default trained with cross-entropy (CE) loss:

\begin{align}
    \mathcal{L}_\text{CE}(\mathbf{x}, \mathbf{y}) = -\frac{1}{C}\sum_{i=1}^{C}(y_{i}\log(x_{i}))
\end{align}

\noindent where $\mathbf{x} = [x_\text{tar}, x_\text{non}, x_\text{spf}]$ stores the output predicted posterior probabilities, $\mathbf{y} = [y_\text{tar}, y_\text{non}, y_\text{spf}]$ is an one-hot vector encoding the ground truth class. $i \in [1,C]$ indexes the vectors. Such a classifier with explicit computation of the three posterior probabilities has been proposed in \cite{joint_opt_sasv2020}, which shares the same paradigm as Fig. \ref{fig:sasv_framework}. However, the proposed classifier accepts the ASV and CM modules as part of the network setup. Apart from that, in \cite{prob_fusion_asvspoof2022} two classifiers were implemented, respectively at score-level (simple fusion) and embedding-level (the concatenated ASV and CM embeddings are input to a DNN, returning binary decisions). However, it also needs ASV and CM embeddings to appear on the input side at the same time.
    
\subsubsection{The decision policy}
This module translates the posterior probabilities or their predecessor representations into the final binary decision. Most of the earlier works place this backend jointly with the classifier, directly outputting the binary decision \cite{yamagishi21_asvspoof, prob_fusion_asvspoof2022, wang2020asvspoof}. As a representative exception, inspired by \cite{eps_framework2014}, the posterior probabilities in \cite{joint_opt_sasv2020} are transformed into scores via logical variant (OR) of false acceptance rate (FAR) and false rejection rate (FRR), with the threshold of the optimization of the classifier being taken into account. Inspired by earlier work on parallel scoring fusion \cite{parallel_cm_max2018} and logistic regression for anti-spoofing \cite{logit_antispoofing2013}, we adopt a simple log-likelihood (LLR) scoring scheme. Denoting the class probabilities as $\theta_\text{tar}, \theta_\text{non}$, and $\theta_\text{spf}$ as in Fig. \ref{fig:baseline_protocol}, we compute the score as below:

\begin{align}
    s = \log\Big(\frac{\theta_\text{tar}}{\alpha \cdot \theta_\text{non} + (1 - \alpha) \cdot \theta_\text{spf}}\Big)
\label{eq:llr_scoring}
\end{align}

\noindent where the parameter $\alpha \in [0,1]$ indicates the relative weights between the non-target and spoof posteriors. Note that since this module is only relevant to the evaluation and fixed after creating the baseline, it is not presented in the figures for the remainder of the paper.

Until this point, we have outlined a DNN-based backend in order to generalize the ASV system against spoofing attacks, with only ASV embeddings as the input. We would like to then investigate how to improve the performance of the backend by taking advantage of spoofing samples we have during the training. Due to the very different task nature and challenges involved in speaker verification and spoofing detection tasks, we primarily focus on domain adaptation (DA) methods and integration of spoof ingredients via multi-task learning, which are detailed in Sections \ref{sec:adaptation} and \ref{sec:learning_spoof}, respectively.

\section{Domain Adaptation for Generalizing ASV Against Spoofing}
\label{sec:adaptation}
Since the spoofing scenario and common speaker verification can be treated as two very different domains, we hypothesize that supervised DA can be helpful to improve the ASV generalizability against spoofing attacks. Therefore, in this section, we present our attempts to optimize the baseline system using DA, which has been successfully applied to other tasks such as speaker adaptation for speech recognition \cite{adaptation2021}.

Since the amount of CM training data we have is relatively small compared to common speech tasks, adapting all weights may not be a proper solution compared to adapting only certain crucial parameters such as BN ones \cite{layertrans2020}. However, it is also well-noted that the network proposed here is relatively small compared to other neural CMs aforementioned \cite{lcnn}. Therefore, we perform DA on the network via two different approaches, depending on whether additional parameters are introduced or not: (1)~\emph{Network-wise adaptation}, where certain learnable parameters that already exist in the layers are updated, such as FC ones ($W_{i}, b_{i}$) and BN ($\mu_{i}, \sigma_{i}$) parameters; (2)~\emph{Structural transform adaptation \cite{structural_adaptation2017}}, where certain components in the network such as activation function or normalizer are parameterized with a transformation matrix $W$ and optionally an optional bias $b$. They are initialized in the network originally as an identity matrix and zero-valued vector respectively, then updated via gradient descent during the adaptation phase, with preferably certain constraints specific to both task nature and practice. 
This approach has been effective in other speech processing tasks such as speech recognition \cite{lhuc2016, adaptation2021}.

\subsection{Network-wise Adaptation}
We first perform network-wise adaptation via fine-tuning \cite{adaptive_finetuning2019} on the backend model. The operations of each of the first two dense layers can be defined as:

\begin{align}
    l(\mathbf{x}) = \beta_{i} + \gamma_{i} \frac{\max((\mathbf{W_{i}}\mathbf{x} + b_{i}), 0) - \mu_{i}}{\sigma_{i}}
\label{eq:dnn_layer_prop}
\end{align}

\noindent which consists of several learnable parameters. We categorize those parameters according to the steps shown in the figure: (1) Affine layer (FC), where the kernel parameters are its weight $W_{i}$ and bias $b_{i}$; (2) BN layer (BatchNorm), where the parameters are the scalar $\gamma_{i}$ and the offset $\beta_{i}$ ($\mu_{i}$ and $\sigma_{i}$ are the running mean and variance of the batch.).

\subsection{Structural Transformation on ReLU}
Our focus then goes to the activation function. We adopt the idea of \cite{lhuc2016} to parameterize the ReLU function. The function is extended into \emph{scaled} ReLU (sReLU) \cite{scaled_relu_vit2022}:

\begin{align}
    \text{sReLU}_{\mathbf{W_{a}},b_{a}}(x) = 
    \begin{cases}
        \mathbf{W_{a}}x + b_{a} & \text{if} x \geq 0 \\
        0 & \text{otherwise}
\end{cases}
\label{eq:srelu}
\end{align}

\noindent where $\mathbf{W_{a}}$ is the scaling matrix. In order to preserve the gradient stability of the adaptation process, we follow two design choices from \cite{adaptation2021}: (1) Unlike with \cite{scaled_relu_vit2022}, we do not parameterize the negative values; (2) $\mathbf{W_{a}}$ is initialized as an identity matrix and restricted to be diagonal. 

Experiments suggested that the bias vector $b_{a}$ harms the performance, so it is not included. If we substitute ReLU with sReLU, Eq. \eqref{eq:dnn_layer_prop} becomes:

\begin{align}
    l(\mathbf{x}) = \beta_{i} + \gamma_{i} \frac{\max(\mathbf{W}_{a}(\mathbf{W}_{i}\mathbf{x} + b_{i}), 0) - \mu_{i}}{\sigma_{i}}
\label{eq:full_dnn_layer_prop}
\end{align}

\noindent where $\mathbf{W}_{a}$ is updated on its own or along with other parameters in the equation.

\begin{figure}[t]
    \centering
    \includegraphics[width=0.9 \linewidth]{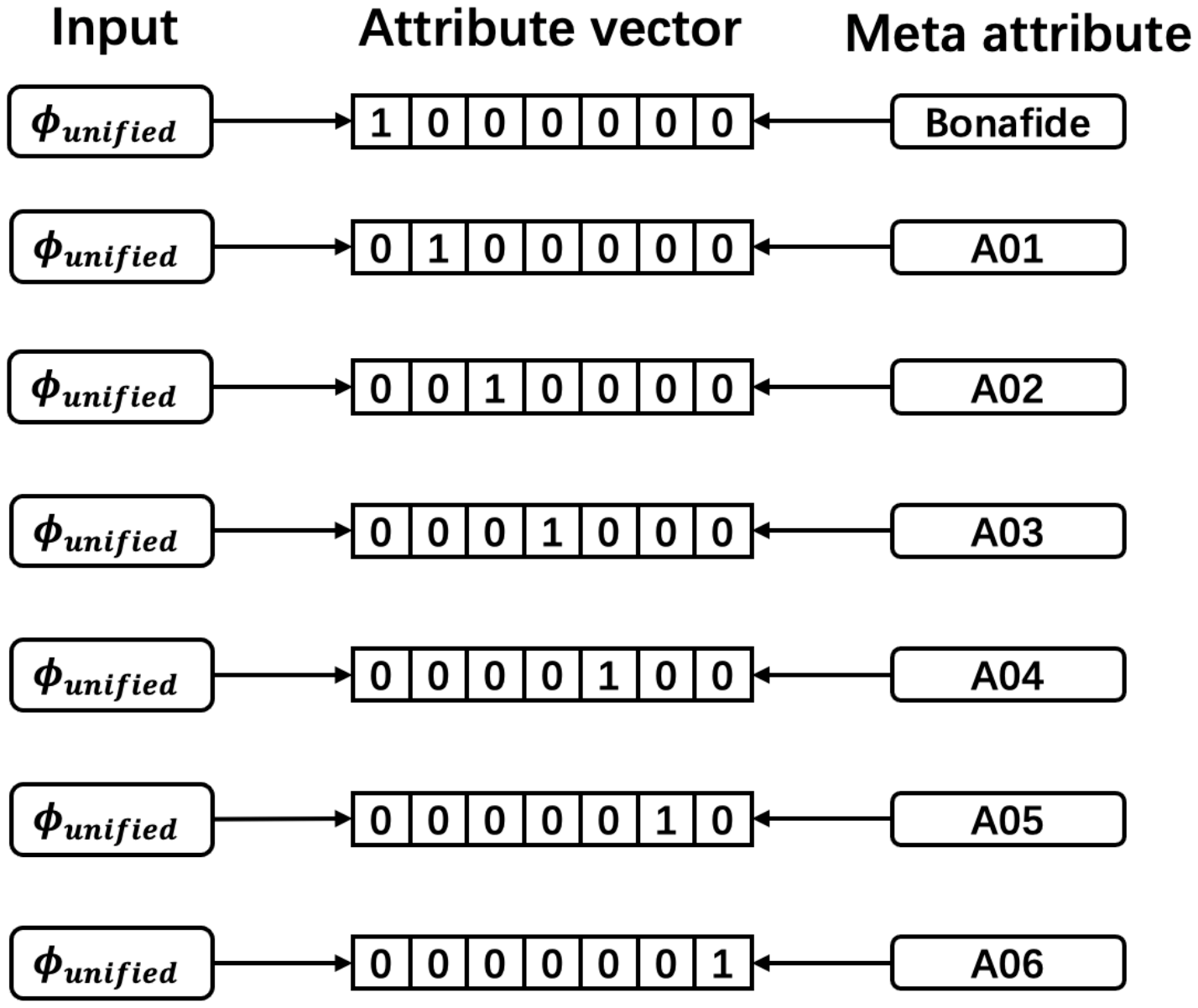}
    \caption{The process of generating the meta attribute vector $\mathbf{\phi}_\text{attr}$ according to the type of attack.}
    \label{fig:meta_attribute}
\end{figure}

\section{Spoof Integration for Generalizing ASV Against Spoofing}
\label{sec:learning_spoof}

The usefulness of CM data to generalize the ASV system has been anticipated in the last section. Building upon this notion, we hypothesize that at the training stage, leveraging the power of a standalone CM in a certain form can bring further improvement to the generalization power of the ASV backend against spoofing attacks.
In this section, we present our approaches towards this objective.
We start by introducing two types of source spoof features as utterance-level vectors, followed by adopting them via a multi-tasking learning framework, with multiple extensions. In such way, we introduce spoof ingredients for training a strong ASV backend without relying on them after training by only utilizing the main classification task.

\subsection{Type of spoof features}
\subsubsection{Synthetic spoof embeddings}
We first utilize a CM binary classifier at the training stage, in order to create embeddings which encode information that is useful for spoof detection. To this end, we use the state-of-the-art AASIST \cite{aasist2022} approach based on RawNet2 \cite{rawnet2_spoofing2021} frontend and graph attention, which uses a roughly 4-second audio waveform as input. The spoof embeddings are then extracted from the fully connected layer prior to the output (binary decision) layer. We follow \cite{prob_fusion_asvspoof2022} to extract 160-dimensional embedding for each utterance. This process is applied to all training data, including VoxCeleb data which is specified via the VoxSRC trials.

\subsubsection{Meta attributes} 
We also exploit information related to the details of the attacks, such as the attack type, vocoder, and waveform generation methods of CM training data \cite{wang2020asvspoof}. Here, we consider the type of attack as an additional attribute vectors in the training data. This process is illustrated in Fig~\ref{fig:meta_attribute}. In particular, each training utterance corresponds to a one-hot vector of $(N+1)$ dimensions, where $N$ is the total number of attacks available in the training data (here, 6), along with the bonafide class.

\begin{figure}[t]
    \centering
    \subfloat[Hard parameter sharing (HPS)\label{fig7:mb}]{
    \includegraphics[trim={2cm 0 2cm 1cm}, width=0.65\linewidth]{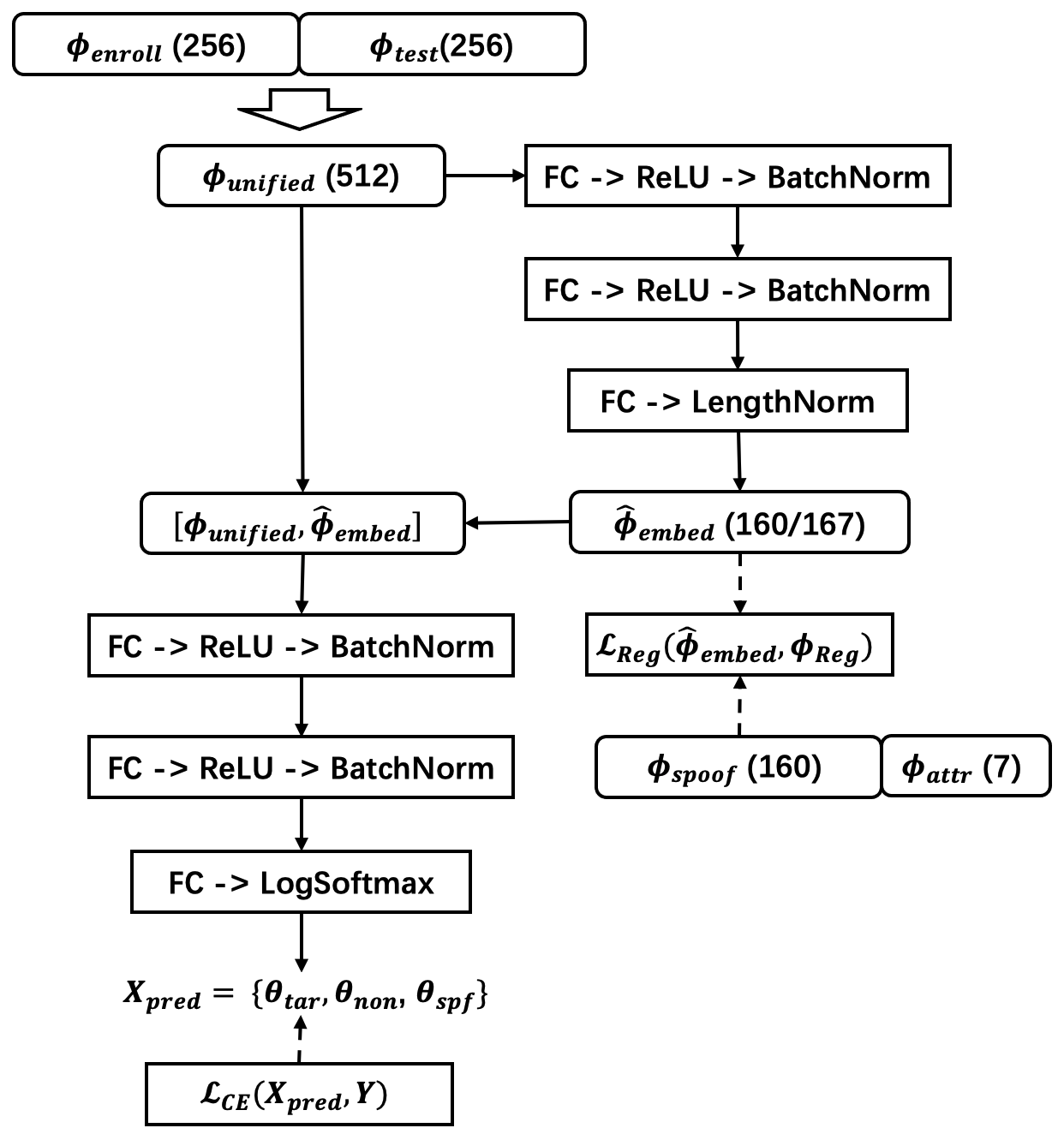} 
    } \par
    \subfloat[Soft parameter sharing (SPS)\label{fig7:mt}]{
    \includegraphics[trim={2cm 0 2cm 1cm}, width=0.625\linewidth]{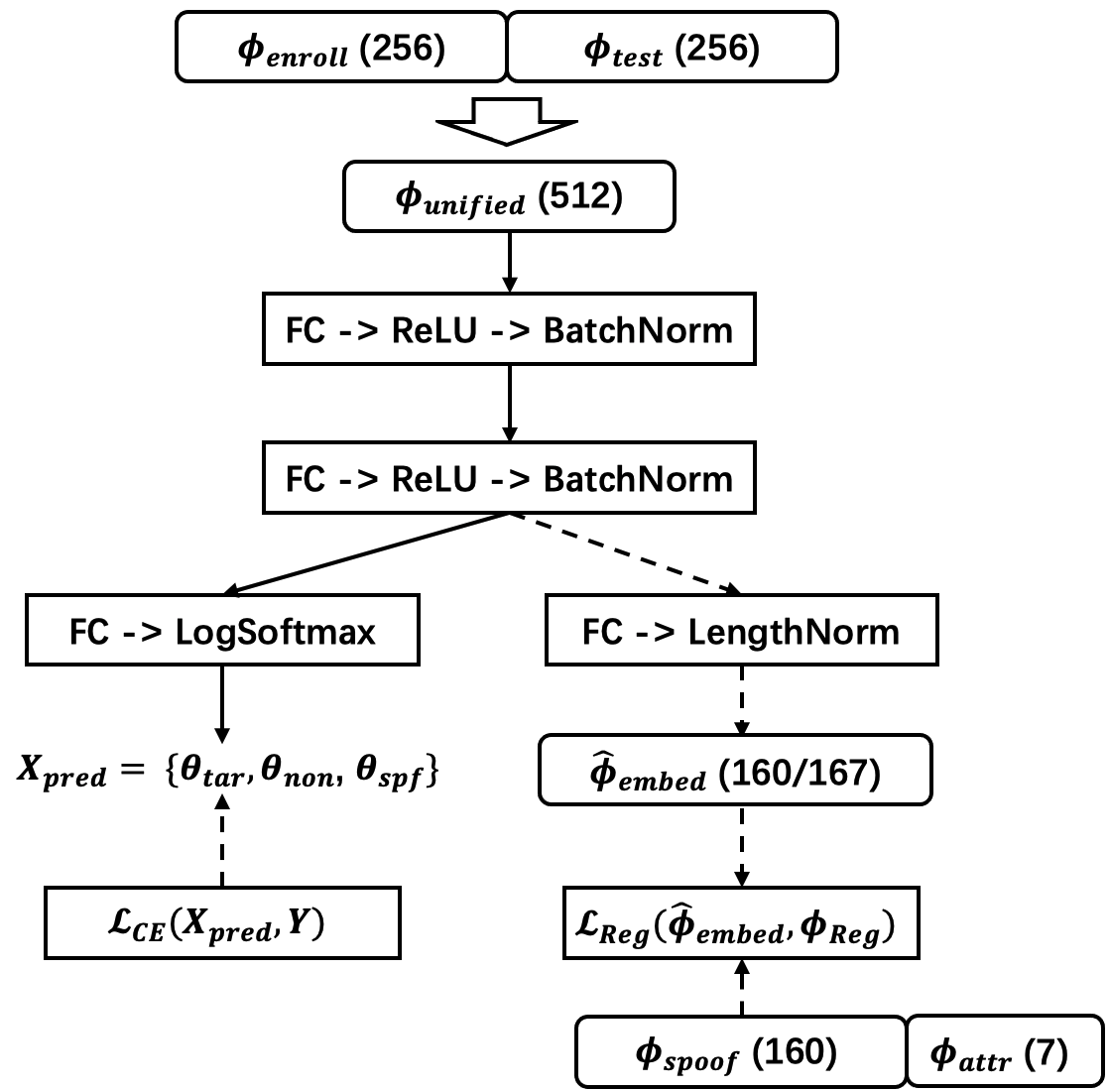} 
    }
    \caption{The multi-task learning schemes for the separate regression branch. The dashed lines indicate steps that are discarded at evaluation. The target distribution of the regression branch $\mathbf{\phi}_\text{reg}$ can be either $\mathbf{\phi}_{\text{spoof}}$ or $[\mathbf{\phi}_{\text{spoof}}, \mathbf{\phi}_{\text{attr}}]$.}
    \label{fig:regression_scheme}
\end{figure}

\begin{figure}[t]
\centering
\subfloat[Hard parameter sharing (HPS-ATTR) with auxiliary classifier\label{fig7:mb_bnf}]{
\includegraphics[width=\linewidth]{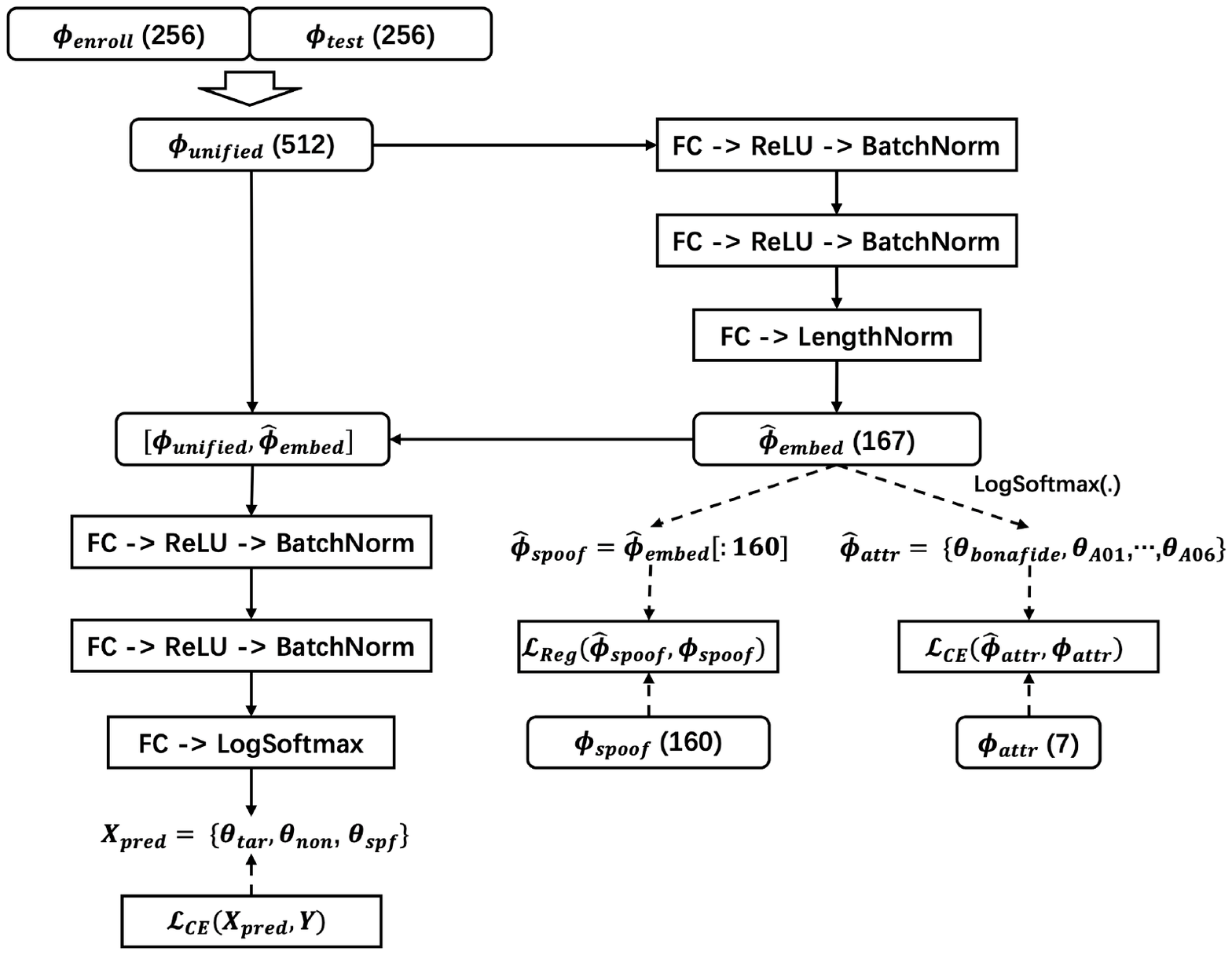} 
} \par
\subfloat[Soft parameter sharing (SPS-ATTR) with auxiliary classifier\label{fig7:mt_bnf}]{
\includegraphics[width=\linewidth]{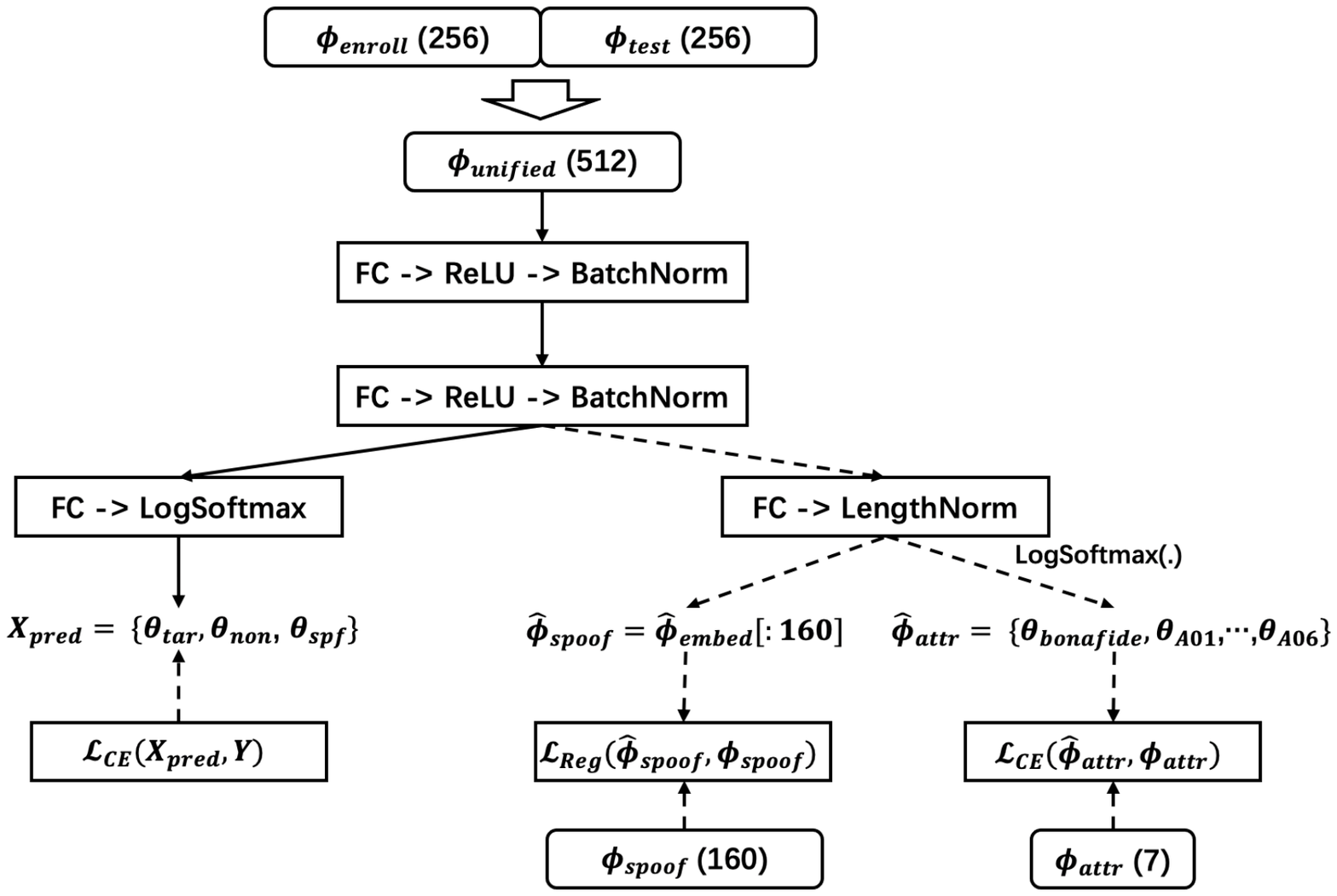} 
}
\caption{The multi-task learning schemes use auxiliary branches with regression and classification with meta attributes as labels ($\mathbf{\phi}_\text{attr}$). The dashed lines indicate steps which are discarded at evaluation.}
\label{fig:regression_bnf_scheme}
\end{figure}

\subsection{Learning Representation of the Spoof Features via Multi-Tasking}
We adopt a multi-task learning paradigm \cite{mtl1997, latent_mtl2019} by creating a separate branch, as shown in Fig~\ref{fig:regression_scheme}. Here, we utilize a separate regression-based learning branch from the main classification branch, which has its learning objective apart from the main cross-entropy objective. For the rest of this section, we denote the cross-entropy loss in the main branch and the regression loss from the auxiliary branch as $\mathcal{L}_\text{CE}$ and $\mathcal{L}_\text{Reg}$, respectively. The total training loss is then expressed as

\begin{align}
    \mathcal{L}_\text{total} = \lambda \times \mathcal{L}_\text{CE} + (1 - \lambda) \times  \mathcal{L}_\text{Reg}
\label{eq:mt_base}
\end{align}

\noindent where $\lambda \in (0,1)$ is an interpolation parameter.

We discuss three design choices under this multi-task learning framework as follows.

\subsubsection{Regression Loss Function}
Perhaps the very first problem is the regression loss function. Denoting $\mathcal{X}_{out}$ as the output prediction from the regression branch and $\mathcal{X}_{tar}$ as the target distribution provided, we primarily employ two loss objectives: (1) \emph{Mean square error} (MSE) \cite{statistical_ml_r2013}: $\mathcal{L}_\text{Reg}(\mathcal{X}_\text{out}, \mathcal{X}_\text{tar}) = {\Vert\mathcal{X}_\text{out} - \mathcal{X}_\text{tar}\Vert}^{2} / N_\text{samples}$. $N_\text{samples}$ denotes the number of samples included; (2) \emph{Cosine similarity} \cite{cosine_similarity}: -$\mathcal{L}_\text{Reg}(\mathcal{X}_\text{out}, \mathcal{X}_\text{tar}) = DOT(\mathcal{X}_\text{out}, \mathcal{X}_\text{tar}) / \mathrm{max}({\Vert\mathcal{X}_\text{out}\Vert}_{2}, {\Vert\mathcal{X}_\text{tar}\Vert}_{2}, \epsilon)$, where $\text{max}$ is the max function, and $\epsilon = 1e^{-8}$ is placed to avoid division by zero. $DOT(.)$ performs the dot product on the two inputs.

\subsubsection{Parameter Sharing Scheme}
While the output and learning objective between the two learning branches are different, the input features and hidden architectures can be identical. Here, we consider two different parameter sharing schemes: soft and hard \cite{mtl_ruder2017, latent_mtl2019}, and as shown in Fig. \ref{fig7:mb} and \ref{fig7:mt}, named \emph{soft parameter sharing} (SPS) and \emph{hard parameter sharing} (HPS), respectively. The key difference between the two architectures is in the way the spoofing information learned by the regression branch is shared. For SPS, the hidden layers between the two branches are shared by the same input. Meanwhile, for HPS, while the two branches do not share any hidden layer parameters, the spoof features learned by the regression branch are appended to the input of the main branch as auxiliary features. For both variants, the classification and regression objectives are jointly optimized. The target distribution for the regression branch is either spoof embedding ($\mathbf{\phi}_{\text{spoof}}$, 160-dimensional) or concatenation of spoof embedding and the attribute ($[\mathbf{\phi}_{\text{spoof}}, \mathbf{\phi}_{\text{attr}}]$, 167-dimensional).

\subsubsection{Learning Spoof Features via Auxiliary Classification}
Moving on from the parameter sharing schemes, another design idea is to incorporate both classification and regression loss into the regression learning branch, thanks to the availability of the attribute vectors. As shown in Fig. \ref{fig7:mb_bnf} and \ref{fig7:mt_bnf}, the output of the final hidden layers of the separated learning branch is fed into two separated sub-branches: on the left-hand side, the regression objective with respect to the spoof embeddings is utilized as above, while on the right-hand side, the posterior probabilities output by the log softmax layer is used to compute another cross-entropy loss, with the spoof attributes as the labels. The target distribution for the regression branch is then only $\mathbf{\phi}_\text{spoof}$, while $\mathbf{\phi}_\text{attr}$ denotes the classification labels for the new branch.

The learned embeddings can be extracted from the left-hand side of the branch as ``attribute" (ATTR) features. Since this incremental design operates only at the output side of the regression branch in the above two sets of variants, the schemes of HPS and SPS schemes remained. We denote the two network architectures as HPS-ATTR and SPS-ATTR, respectively. In both cases, the loss function is then contains three components, with the two in Eq.~\eqref{eq:mt_base} and the additional classification loss via spoof attributes $\mathcal{L}_\text{attr}$. The joint loss function then becomes:

\begin{align}
    \mathcal{L}_\text{total} = \lambda \times \mathcal{L}_\text{CE} + (1 - \lambda) \times ((1 - \gamma) \times \mathcal{L}_\text{Reg} + \gamma \times \mathcal{L}_\text{attr})
\label{eq:mt_bnf_base}
\end{align}

\noindent In the equation above, $\gamma \in (0,1]$ is the additional tuning parameter for the regression branch. We note the tuning parameters above with consideration of the priority of different tasks. Here, the main task corresponds to $\mathcal{L_\text{CE}}$. In such a manner, we hypothesized that the separate classification could better learn underlying information from the meta attributes and assist the learning of the main branch since whichever type of attributes is used, each testing sample is supposed to correspond to only one of the presented attribute variants.

\section{Experimental Setup}
\label{sec:experiments}

\subsection{Data processing and Training}
\label{secsec:data_training}
\subsubsection{Data}
The data we choose is from ASVspoof 2019 LA, which was designed for training the binary CM classifier. We use the utterances from the training partition according to the data partitioning detailed in \cite{wang2020asvspoof}. The training set contains 20 speakers and 2580 bonafide utterances. Each of those 20 speakers correspondingly also has their spoofed counterparts from six different types of attacks (labeled A01-A06). To generate the training samples with pair of enrollment and test segments, we combine each bonafide utterance with both other bonafide utterances. (if the two utterances in the pair belong to the same speaker, the decision is bonafide target, otherwise bonafide non-target) and spoof utterances from the same speaker (spoof target). We denote this dataset as \emph{CM Train}. Additionally, we include VoxCeleb1 data in the training set \cite{voxsrc2019}. In the challenge, two trial lists are presented using the VoxCeleb1 \cite{voxceleb1} dataset: \emph{VoxCeleb1-E} and \emph{VoxCeleb1-H}. We use these trial lists to create the training set with bonafide-target and bonafide-non-target types. The number of training samples for the datasets mentioned above concerning the three classes is shown in Table \ref{tab:trial_info}. For all pre-trained models in the adaptation (described in Section \ref{sec:adaptation}) and systems in this study, we use the union of the three sub-sets, which results in over 2.4 million training samples.

\subsubsection{Speaker model}
We consider the state-of-the-art ECAPA-TDNN \cite{ecapa_tdnn2020} as our ASV model to create speaker embeddings for both enroll and test utterances. We re-implemented ECAPA-TDNN from the original 192-dimensional architecture from \cite{ecapa_tdnn2020} with the embedding dimension being 256. The embeddings are extracted from the first fully-connected layer after the attentive statistics pooling layer. As shown in Fig. \ref{fig:baseline_protocol}, the two embeddings that correspond to the enrolment and test are concatenated as inputs of the backend network.

\subsubsection{Training and model hyper-parameters} 
The backend network is trained via Adam \cite{adam}. The weight decay factor for the optimizer is $1e^{-7}$. The learning rate is scheduled via step-wise decay with a multiplication factor $\gamma_{m}=0.1$, and with an initial value of $0.001$. The data is shuffled and the minibatch size is 128. For all related hyper-parameters including the LLR interpolation one ($\alpha$ in Eq. \eqref{eq:llr_scoring}) and the two weighting parameters for loss propagation ($\lambda$ in Eq. \eqref{eq:mt_base} and \eqref{eq:mt_bnf_base}, $\gamma$ in Eq. \eqref{eq:mt_bnf_base}), are set as $\alpha = 0.95, \lambda = \gamma = 0.5$ unless otherwise specified. In Section \ref{secsec:baseline}, we analyze the impact of those parameters.

\subsection{Evaluation}
We evaluate the systems on the ASVspoof 2019 LA evaluation set, using three different types of EERs that are defined based on their type of the corresponding negative trials, with the bonafide target as positive samples: 1) Joint EER, where both the bonafide non-target and spoof target trials are regarded as negative samples; 2) Bonafide EER, where the bonafide non-target trials are negative samples; 3) Spoof EER, where the spoofed target trials are negative samples. The last two EERs are computed via accordingly-split subsets of the trial list. This set of metrics resembles that applied in \cite{prob_fusion_asvspoof2022, joint_opt_sasv2020}, and \cite{sasv_personal_odyssey2022}.

\begin{table}[ht]
    \centering
    \caption{Details of training data for all the systems presented in this study. We regard the two VoxCeleb subsets from \cite{voxsrc2019} as bonafide data and here they contribute to creating only target and non-target subsets.}
    \begin{tabular}{|c|c|c|c|}
    \hline
        Dataset & Target & Non-target & Spoof \\ \hline
        \emph{CM train} & 330260 & 619576 & 326064 \\ \hline
        \emph{VoxCeleb1-E} & 289921 & 289897 & - \\ \hline
        \emph{VoxCeleb1-H} & 275406 & 275488 & - \\ \hline
    \end{tabular}
    \label{tab:trial_info}
\end{table}
\vspace{-0.3cm}

\begin{table}[ht]
    \centering
    \caption{Experiments on the protocol setup and baseline system. The PLDA baseline results are obtained based on ASV protocols described in ASVspoof 2019 LA \cite{wang2020asvspoof}. We set $\alpha=0.95$ for the last two rows.}
    \begin{tabular}{|c|c|ccc|}
    \hline
        Model & Dataset & Joint & Bonafide & Spoof \\ \hline
        Cosine \cite{cosine_similarity} & - & 25.53 & 1.43 & 33.11 \\ \hline
        PLDA \cite{kaldi_aplda} & - & 26.57 & 1.02 & 35.07 \\ \hline
        MLP & CM train & 16.03 & 20.69 & 13.28  \\ \hline
        MLP & CM train~+~VoxSRC & 12.79 & 5.99 & 15.51 \\ \hline
    \end{tabular}
    \label{tab:baseline_data_exp}
\end{table}
\vspace{-0.3cm}

\begin{figure}[ht]
    \centering
    \includegraphics[trim={2cm 0 2cm 0}, width=0.9\linewidth]{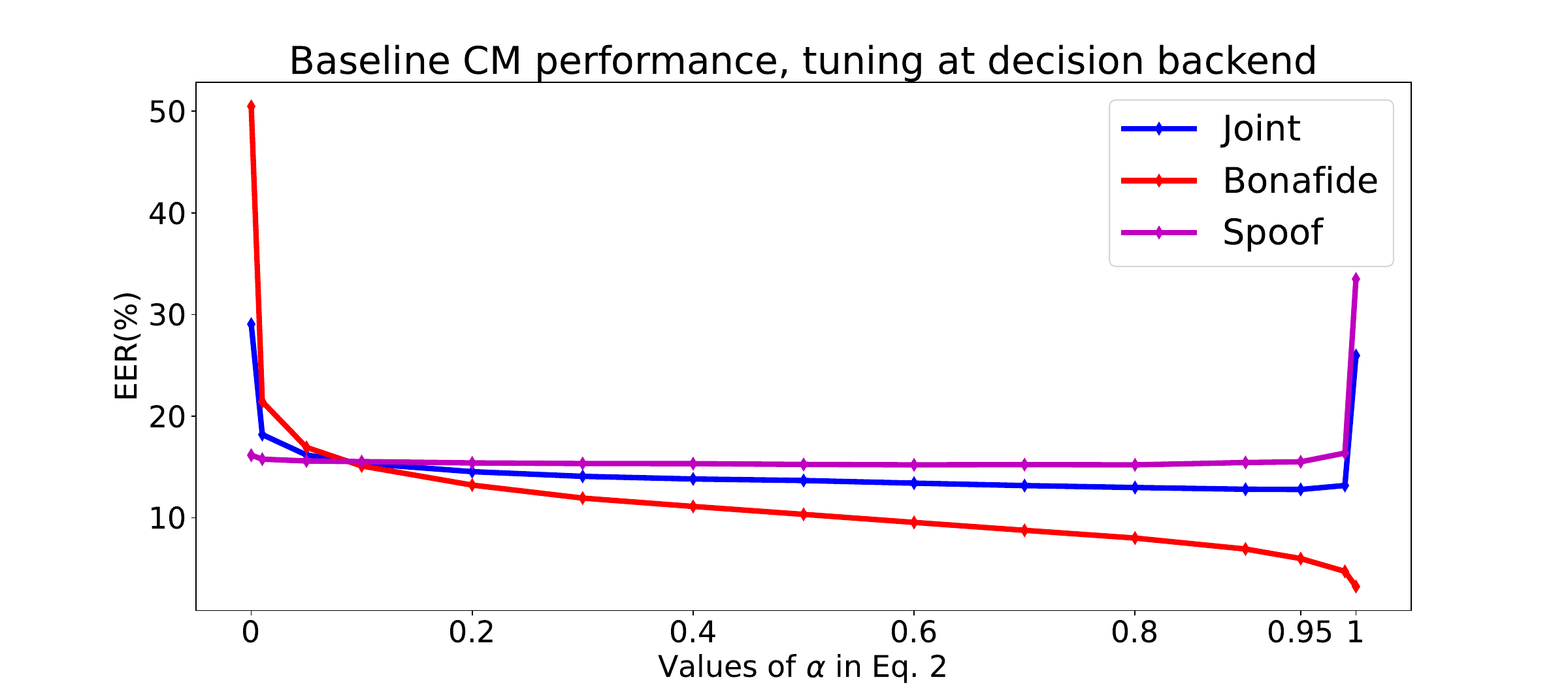}
    \caption{Experiments on tuning $\alpha$ in Eq. \ref{eq:llr_scoring} of the decision backend of the baseline system (last row of Table \ref{tab:baseline_data_exp}). When $\alpha$ equals 0 or 1, the estimation becomes equivalent to that of a standard countermeasure or ASV, respectively. Best viewed in color.}
    \label{fig:baseline_alphas}
\end{figure}

\section{Results and Analysis}
\label{sec:results}

\subsection{Baseline Experiments}
\label{secsec:baseline}
We first conduct a set of initial experiments related to the baseline SASV as described in Section \ref{sec:sgsv_def}. From these experiments, we select the tuning parameters $\alpha$ of our architecture and we optimize the training data requirement. In addition, we also compare our methods with related state-of-the-art systems.

First, we experimentally optimize the value of $\alpha$ in Eq.~\eqref{eq:llr_scoring}. Results are illustrated in Fig.~\ref{fig:baseline_alphas}. As the value of $\alpha$ increases, while the baseline performance in terms of bonafide EER improves, on the spoofing set the improvement is rather marginal, and is notably worsened when $\alpha$ goes from 0.95 to 1. The best joint EER is obtained at $\alpha=0.95$ and we use this parameter for the remaining set of experiments.

Second, we address the potential benefit of including the VoxCeleb data. The results are shown in Table \ref{tab:baseline_data_exp} along with the baselines obtained by cosine scoring and PLDA, following the ASV system described in \cite{wang2020asvspoof}. The PLDA system is trained on VoxCeleb1 \cite{voxceleb1}, which provides the source data of the two VoxSRC trials. There are two interesting findings. First, compared to the conventional backends, the MLP backend results in notable improvements on joint and spoof EERs, with a certain degree of performance cost on the bonafide subset. This may be because compared to backends like PLDA where unsupervised adaptation is performed, the DNN backend is trained in a supervised fashion, taking advantage of the designated spoof labels (corresponding to the three classes). Second, including additional bonafide data leads to further improvements on joint and bonafide EERs while with marginal performance cost on spoof EER, which is expected due to the potential mismatch between the bonafide and spoof scenario.

\begin{table}[th]
    \centering
    \caption{Experiments on the adaptation. The adapted components covered in the experiments include the entire network (Network), affine weights and biases (FC), batch normalization parameters (BN), or sReLU weights (ReLU). The baseline is trained without any adaptation or addition.}
    \begin{tabular}{|c|c|ccc|}
    \hline
        Adaptation & Addition & Joint & Bonafide & Spoof \\ \hline
        - & - & 12.79 & \textbf{5.99} & 15.51 \\ \hline
        Network & - & 11.19 & 11.97 & 10.73 \\ \hline
        FC & - & 11.45 & 12.66 & \textbf{10.71} \\ \hline
        BN & - & \textbf{10.07} & 6.95 & 11.66 \\ \hline
        ReLU & sReLU & 11.82 & 7.56 & 14.21 \\ \hline
        Network & sReLU & 11.94 & 15.22 & 10.86 \\ \hline
    \end{tabular}
    \label{tab:adaptation_exp}
\end{table}

\begin{figure}[th]
    \centering
    \includegraphics[trim={2cm 0 2cm 1cm},width=0.9\linewidth]{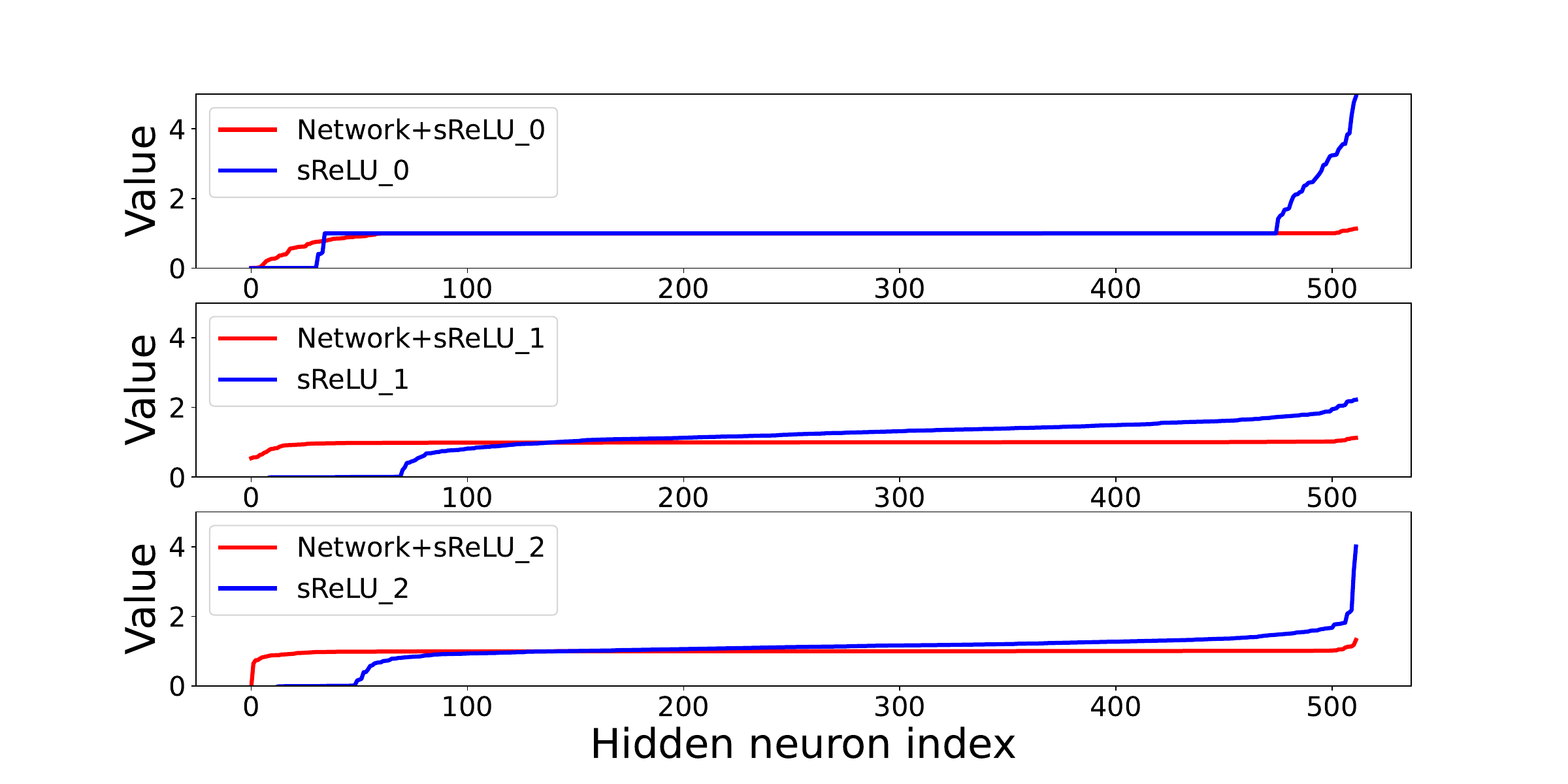}
    \caption{Learned activation representations at each of the three layers in the two systems with sReLU addition. Best viewed in color.}
    \label{fig:lhuc_representation}
\end{figure}

\begin{table}[t]
    \centering
    \caption{Experiments on spoof methods via multi-task learning, with varying regression loss functions (Reg. Loss), parameter sharing schemes (PSS), and target representations for regression (Target).}
    \begin{tabular}{|c|c|c|ccc|}
    \hline
        Reg. Loss & PSS & Target & Joint & Bonafide & Spoof \\ \hline
        Cosine & SPS & $\mathbf{\phi}_{\text{spoof}}$ & \textbf{9.42} & 11.96 & \textbf{8.08} \\ \hline
        MSE & SPS & $\mathbf{\phi}_{\text{spoof}}$ & 10.13 & 10.47 & 9.89 \\ \hline
        Cosine & HPS & $\mathbf{\phi}_{\text{spoof}}$ & 9.83 & 10.89 & 9.46 \\ \hline
        MSE & HPS & $\mathbf{\phi}_{\text{spoof}}$ & 10.52 & \textbf{9.66} & 11.02 \\ \hline
        Cosine & SPS & $[\mathbf{\phi}_{\text{spoof}}, \mathbf{\phi}_{\text{attr}}]$ & 10.39 & 11.77 & 9.57 \\ \hline
        MSE & SPS & $[\mathbf{\phi}_{\text{spoof}}, \mathbf{\phi}_{\text{attr}}]$ & 11.42 & 12.66 & 10.65 \\ \hline
        Cosine & HPS & $[\mathbf{\phi}_{\text{spoof}}, \mathbf{\phi}_{\text{attr}}]$ & 10.24 & 9.72 & 10.48 \\ \hline
        MSE & HPS & $[\mathbf{\phi}_{\text{spoof}}, \mathbf{\phi}_{\text{attr}}]$ & 11.12 & 12.74 & 10.26 \\ \hline
    \end{tabular}
    \label{tab:adaptation_exp_spoof_attr}
\end{table}

\begin{figure}[th]
    \centering
    \includegraphics[trim={2cm 0 2cm 1cm}, width=\linewidth]{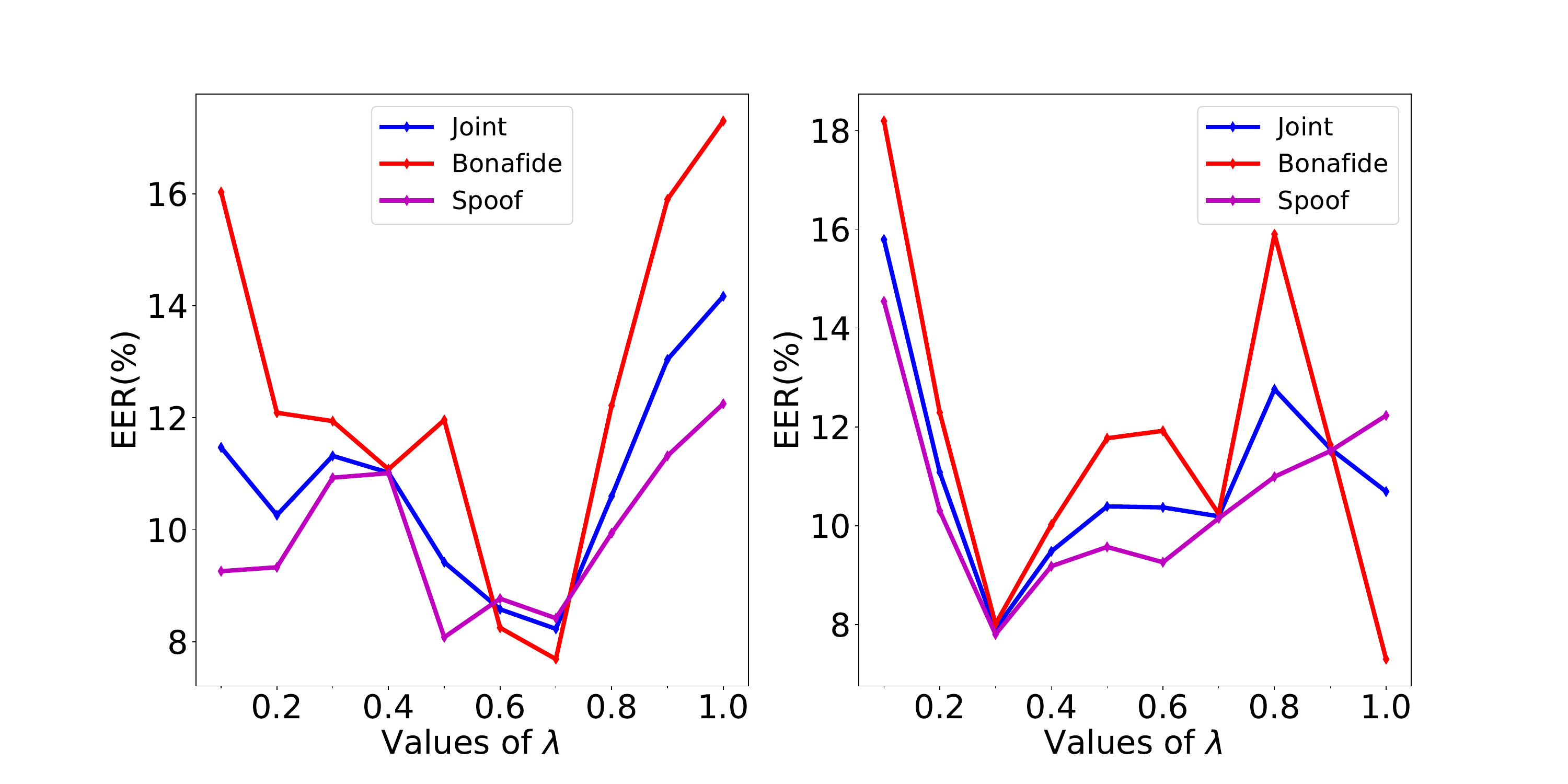}
    \caption{The tuning experiment on $\lambda$ in Eq.~\ref{eq:mt_base}. The target distribution for the regression branch is either $\boldsymbol{\phi}_\text{spoof}$ (left) or $[\boldsymbol{\phi}_\text{spoof}, \boldsymbol{\phi}_\text{attr}]$ (right). The loss function for the regression branch is cosine similarity. The later sharing scheme is SPS. Best viewed in color.}
    \label{fig:tuning_lambdas_eq7}
\end{figure}

\begin{table}[th]
    \centering
    \caption{Experiments on spoof methods via multi-task learning with a bottle-neck sharing scheme.  with varying regression loss functions (Reg. Loss), and with parameter sharing schemes (PSS).}
    \begin{tabular}{|c|c|ccc|}
    \hline
        Reg. Loss & PSS & Joint & Bonafide & Spoof \\ \hline
        Cosine & SPS-ATTR & 9.25 & 11.81 & \textbf{7.78} \\ \hline
        MSE & SPS-ATTR & 18.36 & 21.88 & 16.15 \\ \hline
        Cosine & HPS-ATTR & \textbf{8.62} & \textbf{8.54} & 8.64 \\ \hline
        MSE & HPS-ATTR & 17.73 & 11.53 & 20.60 \\ \hline
    \end{tabular}
    \label{tab:adaptation_exp_spoof_attr_bnf}
\end{table}

\begin{figure}[th]
    \centering
    \includegraphics[trim={2cm 0 2cm 1cm}, width=\linewidth]{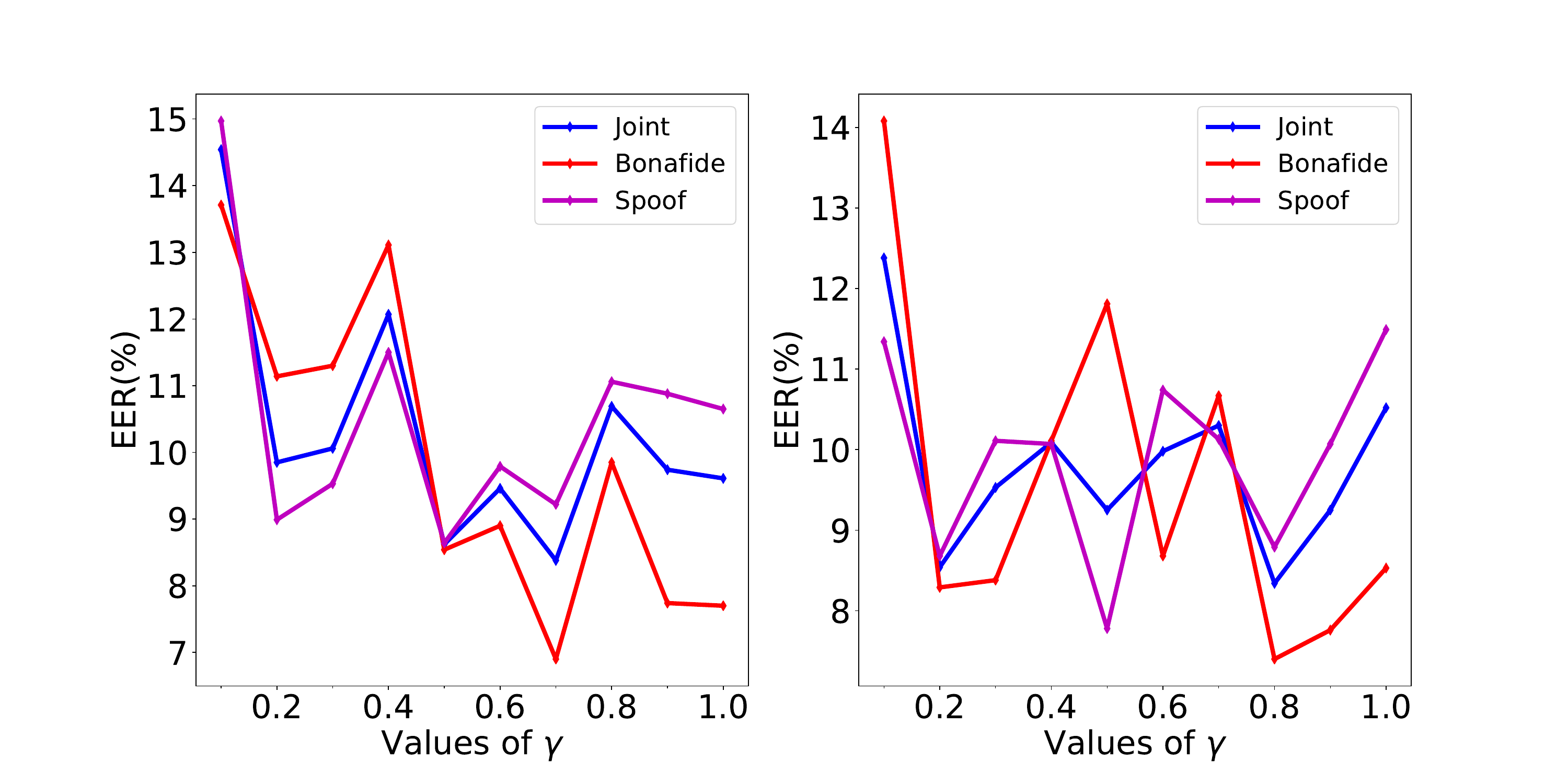}
    \caption{The tuning experiment on $\gamma$ in Eq.~\eqref{eq:mt_bnf_base}. The parameter sharing scheme here is either SPS-ATTR (left) or HPS-ATTR (right). The loss function for the regression branch is cosine similarity. The interpolation parameter $\lambda=0.3$. Best viewed in color.}
    \label{fig:tuning_gammas_eq8_sls}
\end{figure}

\subsection{Experiments with Adaptation}
We perform the next set of experiments related to the adaptation of the SASV models. To evaluate the importance of each of these parameters, in our experiments, we ablate the trainable parameters one by one, along with an experiment where we update all of them. 

We present the results in Table~\ref{tab:adaptation_exp}. The best joint EER is obtained by adapting the BN parameters in Eq.~\eqref{eq:dnn_layer_prop}, outperforming the baseline by relatively 21.3\%. The baseline shows its best performance on bonafide utterances, which indicates that adaptation using CM data may degrade the performance on the bonafide scenarios. This might be because CM audio is already used to train in the first place, and fine-tuning only includes redundant information regarding the bonafide side. Meanwhile, all adapted systems outperform the baseline in terms of spoof EER. Adapting the weights and biases in the affine operation of the layer results in the best performance on the spoofed set, outperforming the baseline by relatively 30.9\%. The bonafide performance of adapting the weights and biases, however, is the second worst among all systems, potentially indicating a trade-off between bonafide and spoofed sets. A similar trade-off is noted at the baseline, which obtains the lowest bonafide EER and highest spoof EER.

Regarding the two systems with sReLU, both outperform the baseline in terms of joint and spoof EERs, implying the efficacy of LHUC for spoofing tasks. However, neither of them produces the lowest EER among any of the subsets. Adapting all parameters in the network with the inclusion of the sReLU results in worse bonafide performance and slightly higher joint EER compared to adapting only parameters in the network without, which indicates possible overfit and thus has less generalizability for unseen attacks. 

To explore the cause of the detrimental effect on bonafide performance by adapting network parameters along with sReLU, we analyze the learned representation of the activation functions. The learned $\mathbf{W}_{a}$ with respect to the hidden neuron index for the two layers in the DNN are shown in Fig.~\ref{fig:lhuc_representation}. For all three layers, the value of learned sReLU when adapting only $\mathbf{W}_{a}$ are more polarized at lower and higher index regions, being a closer match to ReLU values. This suggests that the activation may overfit when jointly optimized with the other network parameters, exacerbating the bonafide performance.

\subsection{Spoof Integration}
There are three main variation factors in the spoof integration design in total, as mentioned in Section~\ref{sec:sgsv_def}, namely, the loss function for the regression network, the parameter sharing scheme, and the target representation. Tables~\ref{tab:adaptation_exp_spoof_attr} and~\ref{tab:adaptation_exp_spoof_attr_bnf} show the results of different combinations of these factors.
First, in terms of the regression loss function, cosine similarity outperforms the MSE in all cases in terms of joint EER, with either a comparable or better performance on spoof EER. This might be due to the fact that the pre-trained ASV model, ECAPA-TDNN, is optimized specifically to cosine similarity measures due to its additive angular softmax loss function \cite{aam_softmax}. As a result, acquiring cosine similarity corresponds to a more effective resemblance between source and target distributions, compared to simple MSE.
Second, while there is no distinguishable difference between the SPS and HPS in terms of performance, SPS with cosine similarity and $X_\text{spoof}$ as the target reached the best joint and spoof EERs among all systems listed here. Switching SPS from HPS results in worse performance in all three metrics, while additionally switching to MSE as a loss function results in the best bonafide EER, with a significant performance cost on the other two metrics. 
Finally, comparing the first and second half of the table, simply appending $X_\text{attr}$ leads to worse performance. This indicates that by integrating $X_\text{attr}$ via simple concatenation to create spoof distributions, it may become the additional noise, as it is not able to express its information and usefulness.

It is then natural to tune the value of $\lambda \in (0,1]$. We perform experiments by experimentally varying the parameter for two cases, depending on whether $X_\text{attr}$ is appended. The results are illustrated in Fig.~\ref{fig:tuning_lambdas_eq7}. With only $X_\text{spoof}$, the best bonafide and joint EERs are obtained by setting $\lambda=0.7$, while the default setting is shown Table~\ref{tab:adaptation_exp_spoof_attr}, where $\lambda=0.5$ comes with the lowest spoof EER. When $\lambda>0.7$, all three EERs grow noticeably higher from their optimal, which indicates the detrimental effect of biasing toward the regression branch on both scenarios. With $[X_\text{spoof}, X_\text{attr}]$, the best performance in terms of joint and spoof EERs are obtained at $\lambda=0.3$, and the best bonafide EERs are observed at $\lambda=1.0$. We subsequently use $X_\text{attr}$ as the target with $\lambda=0.3$.

To further investigate whether the $X_\text{attr}$ could be unveiled and effectively utilized, we utilize a separate classification along with the regression branch using $X_\text{spoof}$. The results are shown in Table~\ref{tab:adaptation_exp_spoof_attr_bnf}. We see that while MSE as a regression loss function is basically not useful here, cosine similarity fits well with both parameter sharing schemes with a bottleneck feature design, taking good advantage of $X_\text{attr}$. Cosine loss with SPS-ATTR reached the best spoof EER, outperforming cosine loss with SPS and either $X_\text{spoof}$ or $[\mathbf{\phi}_{\text{spoof}}, \mathbf{\phi}_{\text{attr}}]$ on all three EERs. It outperforms the baseline on spoof EER by relatively 49.8\%. The same is true for cosine loss with HPS-ATTR, which returns the best joint and bonafide EERs across all proposed systems, and whose joint EER is better than the baseline by 32.6\% relatively. While the usefulness of $X_\text{attr}$ as the classification label is revealed by those advantageous numbers, they also suggest that the 7-dimensional vector needs to be separately utilized, instead of simply appending to $\mathbf{\phi}_{\text{spoof}}$. In such cases, thanks to the separate classification branch, the information of meta attributes can be more properly learned by the network. 

We further tune the value of $\gamma \in (0,1]$ which was set at 0.5 for the initial experiments in Table~\ref{tab:adaptation_exp_spoof_attr_bnf}. The results are shown in Fig.~\ref{fig:tuning_gammas_eq8_sls}. For SPS-ATTR, it can be seen that the performance of the system with respect to the varying $\gamma$ is generally consistent with three EERs. Best joint and bonafide EERs are observed at $\gamma=0.3$, while the setup addressed in Table~\ref{tab:adaptation_exp_spoof_attr_bnf} pertains its best spoof EER, where $\gamma=0.5$. For HPS-ATTR, while the trend of system performance is no longer consistent across different evaluation conditions, $\gamma=0.5$ holds its optimal spoof EER. Therefore, for the next subsequent ablation studies, we set $\lambda=0.3$ and $\gamma=0.5$.

\begin{table}[t]
    \centering
    \caption{Experiments on label smoothing. The loss function for regression here is cosine similarity. Here the two parameters in Eq.~\eqref{eq:mt_bnf_base} are $\lambda=0.3$ and $\gamma=0.5$.}
    \begin{tabular}{|c|c|ccc|}
    \hline
        PSS & $\epsilon$ & Joint & Bonafide & Spoof \\ \hline
        \multirow{3}{5em}{SPS-ATTR} & 0.0 & 9.25 & 11.81 & \textbf{7.78} \\ \cline{2-5}
        & 0.5 & 9.55 & 8.69 & 10.11 \\ \cline{2-5}
        & 1.0 & 8.88 & 9.53 & 8.52 \\ \hline
        \multirow{3}{5em}{HPS-ATTR} & 0.0 & \textbf{8.62} & \textbf{8.54} & 8.64 \\ \cline{2-5}
        & 0.5 & 10.69 & 10.63 & 10.73 \\ \cline{2-5}
        & 1.0 & 9.70 & 8.58 & 10.17 \\ \hline
    \end{tabular}
    \label{tab:label_smoothing}
\end{table}

\begin{table}[ht]
    \centering
    \caption{Experiments on different attributes from the metadata. Here, the loss function for regression here is cosine similarity and the parameter sharing scheme here is SPS-ATTR. The brackets show the dimension of the attribute vector. Here the two parameters in Eq.~\eqref{eq:mt_bnf_base} are $\lambda=0.3$ and $\gamma=0.5$. No label smoothing is applied.}
    \begin{tabular}{|c|ccc|}
    \hline
         Type of attr. (dim.) & Joint & Bonafide & Spoof \\ \hline
         Spoofing attack (7) & 9.25 & 11.81 & \textbf{7.78}\\ \hline
         Vocoder (10) & 9.53 & 9.72 & 9.35 \\ \hline
         Synthesizer (12) & \textbf{8.16} & \textbf{8.21} & 8.12 \\ \hline
         Waveform Generator (10) & 11.97 & 14.25 & 10.73 \\ \hline
    \end{tabular}
    \label{tab:different_spoof_attr}
\end{table}

\section{Ablation Study}
In the last section, we determine the effectiveness of the attribute labels, namely the type of spoofing attack derived from the training set. We therefore investigate label smoothing and different types of attributes that can be derived from the data.

\subsection{Label Smoothing on BNF Loss}
A remaining problem according to the observation above is the imbalanced distribution of the data between classes. This problem can be particularly a concern for attribute learning which is a classification task with one-hot labels. Therefore, it is reasonable to attempt to determine whether label smoothing regularization \cite{label_smoothing} on the attribute classification branch is beneficial. The regularizer operates on the label distribution via simple interpolation, as expressed below:

\begin{align}
    q'(k|x) = (1-\epsilon)q(k|x) + \epsilon u(k)
\label{eq:label_smoothing}
\end{align}

\noindent where $y$ is the ground-truth label, $k \in {1,...,K}$ is the label index, $q(k|x)=\delta_{k,y}$ is the one-hot label distribution, and $\epsilon \in [0, 1]$ is the smoothing weight parameter. In this study, we test three representative values of $\epsilon$: 0.0, 0.5, and 1.0. $u(k) = 1/K$ follows the original definition from \cite{label_smoothing}. Note that $\epsilon=0.0$ means no label smoothing is applied.

Results are shown in Table~\ref{tab:label_smoothing}, covering both SPS-ATTR and HPS-ATTR schemes. For the SPS-ATTR scheme, setting $\epsilon$ to a higher value can lead to better bonafide performance, with the cost of higher spoof EERs. This expresses the favor of the label smoothing method to bonafide ASV. Interestingly, when $\epsilon=1.0$, the best joint EER in SPS-ATTR is reached, besting $\epsilon=0.0$ by relatively 4\% on joint EER and 19.3\% on bonafide EER. While this indicates the usefulness of label smoothing, it also implies that the attribute labels are useful only for the system to be robust to spoofing attacks, with a cost on bonafide ASV performance. The problem of attribute labels may be alleviated by the HPS-ATTR sharing scheme, where we directly attach the learned spoof embeddings to the main classification branch. The best performance on all three metrics under HPS-ATTR is reached with $\epsilon=0.0$, but its spoofing EER is not optimal in this study.

\subsection{The Type of Attribute Labels}
So far in this section, we have used types of attacks (bonafide, A01-A06) as the attribute labels for the attribute classification loss. Since the efficacy of those attribute labels has been observed, it is natural to attempt to determine if other types of meta attributes can be useful for the system. Fortunately, in the case of this study, one can derive various types of attribute labels from the dataset description table in \cite{wang2020asvspoof}\footnote{Interested readers may check Table 1 and Fig 12 of the paper.}. In this paper, we draw three alternative attributes from the dataset description: type of vocoder, type of synthesizer, and waveform generator used. Table~\ref{tab:different_spoof_attr} indicates the number of attribute classes for each case, wherein $D+1$, where $D$ is the number of attribute classes and the extra dimension corresponds to the bonafide.

Results using different attribute labels are presented in Table~\ref{tab:different_spoof_attr}. While the type of spoofing attacks maintains their best spoof EER across all other attempts, the best joint and bonafide EERs among them are obtained by using the type of synthesizer, outperforming the original baseline on joint EERs by relatively 36.2\%. This indicates that type of synthesizer may be more crucial than other attributes for proper generalization of spoof-aware ASV system. Such non-triviality is leveraged to improve the generalizability of the network, with a slight cost on spoofing robustness. Such a cost is reflected by the 4.3\% relatively higher spoof EER than the one that acquires the type of spoofing attacks. The advantageous bonafide performance returned by acquiring the type of synthesizer may be due to the fact that compared to other components such as vocoder or waveform generator, the synthesizer may encode more direct speaker information. This is more likely in the case where ASVspoof LA data is used for training because there are several synthetic algorithms where the synthesizer is designed with speaker adaptation \cite{wang2020asvspoof}.

\subsection{Comparison with Existing SASV Related Studies}
\noindent Finally, we provide a comparison of our approach with earlier existing works on multiple aspects. Both of the two baselines for the SASV challenge \cite{sasv2022}, which acquire the inputs from standalone ASV \cite{ecapa_tdnn2020} and CM \cite{aasist2022} modules, hold strong spoof detection performance. Our system outperforms the single ASV system (ECAPA-TDNN) in terms of spoof EER, and single CM system (AASIST), along with the two SASV baselines (SASV-B\{1,2\}), in terms of bonafide EER. Meanwhile, in \cite{sasv_hyu2022}, where the ASV and CM embeddings are integrated with a subtle integration module based on a gating mechanism, the performance of the system on all embeddings has been improved from the two SASV baselines. 
Similar analogy can be observed for \cite{prob_fusion_asvspoof2022} and \cite{sasv_multilevel2022}. 
While our proposed simple framework improves the performance over both standalone modules without CM and fusion at the authentication stage, future work may take inspiration from the integration module as extending the simple DNN-based backend with multi-task training strategies, especially for the earlier shared layers, in order to let the network more effectively learn the information provided by the spoof features at training stage.

\begin{table}[t]
    \scriptsize
    \centering
    \caption{Comparison with existing SASV related studies in terms of EERs, including standalone ASV and CM systems and the two baseline systems of SASV challenge \cite{sasv2022}. The first two rows correspond to single ASV/CM systems. ``Need CM" indicates the necessity of a standalone CM module at the authentication stage. ``Fusion level" presents the way ASV and CM information are integrated when the CM module is present at the authentication stage. The comparisons presented here are restricted to systems that employ at most one CM and one ASV subsystem.}
    \begin{tabular}{|c|ccc|c|c|}
    \hline
         & Joint & Bonafide & Spoof & Need CM & Fusion level  \\ \hline
         ECAPA-TDNN \cite{ecapa_tdnn2020} & 23.83 & 1.63 & 30.75 & - & - \\
         AASIST \cite{aasist2022} & 24.38 & 49.24 & 0.67 & - & - \\ \hline
        SASV-B1 \cite{sasv2022} & 19.31 & 35.32 & 0.67 & Yes & score \\
        SASV-B2 \cite{sasv2022} & 6.37 & 11.48 & 0.78 & Yes & embedding \\ \hline
        \cite{prob_fusion_asvspoof2022} & 1.53 & 1.94 & 0.80 & Yes & embedding \\
        \cite{sasv_hyu2022} & 0.54 & 0.38 & 0.47 & Yes & embedding \\ 
        \cite{sasv_multilevel2022} & 0.89 & 1.01 & 0.71 & Yes & embedding \\ \hline
        Ours & 8.62 & 8.54 & 8.64 & No & - \\ \hline
    \end{tabular}
    \label{tab:comparison_with_existing_works}
\end{table}

\section{Limitations and Implications}
While we are hopeful that the introduction of G-SASV will bring new insights and solutions to both ASV and anti-spoofing research, we are aware that this study has several limitations and implications for further study.

\subsubsection{The problem of lacking spoofed data}
While observed in Table~\ref{tab:baseline_data_exp} that additional bonafide training examples improves the performance by a large margin, as can be seen in Table~\ref{tab:trial_info}, the number of spoofing trials stays much lower than the number of bonafide trials ($\sim$320k vs. $\sim$2.08M). Meanwhile, in the context of this study, it is hard to generate more spoof training examples. Future work may focus on the effective generation of spoof training examples by introducing more advanced synthetic techniques. Alternatively, incorporating better training strategies under imbalanced data can be considered.

\subsubsection{Data-specific properties of the spoofing data}
The problem of G-SASV, along with other problems listed in Section~\ref{sec:sgsv_def}, may have been tackled from a purely statistical perspective thus far. The additional information incorporated has mainly been pre-defined categorical data, which may not always be practical in real-world scenarios. Future work may focus on discovering and making good use of the underlying data-specific unique physical information such as acoustic properties, artifacts, and spectral representations. Such information has been addressed from the perspective of spoofing detection \cite{partialspoof2022} and shall be applied and extended for generalizing ASV against spoofing.

\subsubsection{The underlying trade-off between bonafide and spoofing scenarios}
From the results in earlier sections, including the pilot experiments on the baseline system, there is a trade-off between bonafide and spoof EERs --- by integrating techniques that bring improvements on spoof EER, the system mostly degrades performance on the bonafide scenario. Exceptions can be found in Table~\ref{tab:baseline_data_exp} and Table~\ref{tab:adaptation_exp_spoof_attr_bnf}, but in those cases, the joint EER is substantially improved (relatively around 50\%) along with other evaluation conditions. However, so far, such a trade-off has only been unveiled from a statistical perspective. Future work may focus on investigating the naturalness trade-off and why it happened, which may incentivize advanced anti-spoofing techniques and in-domain spoofing data collection.

\subsubsection{Optimizing the frontend}
The focus of our work has been to explore the potential of \emph{backend} optimizations to address spoof-aware speaker verification. Even if our results indicates substantial improvement over barebone ASV systems, the performance also depends on the \emph{frontend}. Given the fundamental differences between ASV and anti-spoofing tasks, the selection of acoustic features (here, mel filterbank) and the embedding extractor (here, ECAPA-TDNN) may emerge as bottleneck factors affecting performance. Therefore, despite the promising outcomes achieved in our backend optimizations, future research can focus on optimizing the early-trained frontend speaker embedding extractor, whether joint optimization with the backend classifier.

\section{Conclusion}
\label{sec:conclusion}
In this paper, we first have revisited the different tasks involved in the problem of speaker verification and anti-spoofing. Then, we have introduced the task of generalizing speaker verification systems (G-SASV) against spoofing attacks. We have described the task with baseline systems, decision policy, and evaluation metrics. The proposed DNN-based backend is built upon the speaker embeddings from the state-of-the-art pre-trained ASV model. We have then applied domain adaptation and multi-task learning schemes to model an effective embedding space. These methods take advantage of countermeasure data at training at either the data level or the embedding level, without the reliance on a separate countermeasure module for evaluation/authentication. We have improved the performance from the proposed baseline by a maximum of relatively 36.2\% and 49.8\% on joint and spoof EER, respectively, with comparisons to earlier systems. Future work may focus on architectural improvement from the proposed DNN, tackling the problem of unbalanced data, and bridging the statistical findings with the physical properties underlying the spoofed audios.

\section*{Acknowledgment}
The work has been partially supported by the Academy of Finland (Decision No. 349605, project ``SPEECHFAKES'') and Agency of Science, Technology and Research (A$^\star$STAR), Singapore, through its CRF Core Project Scheme (Project No. CR-2021-005).

\ifCLASSOPTIONcaptionsoff
  \newpage
\fi

\bibliographystyle{IEEEtran}
\bibliography{mybib}

\end{document}